\documentclass[a4paper,11pt]{article}
\usepackage[dvipdfmx]{graphicx}
\usepackage{graphicx}

\usepackage{amsmath}        
\usepackage{amssymb}        
\usepackage{cite}
\usepackage{cancel}
\usepackage{fancybox}
\usepackage{color}
\usepackage{ulem}
\usepackage{pifont}
\usepackage{colortbl}
\usepackage{multicol}
\usepackage{lscape}
\usepackage{colortbl}
\usepackage{multicol}
\usepackage[all,knot]{xy}
\usepackage{longtable}
\usepackage{comment}
\usepackage{bm}
\usepackage{braket}
\usepackage{hhline}
\usepackage{slashed}

\def\tr{\mathop{\text{tr}}}

\def\sqr#1#2{{\vcenter{\hrule height.#2pt
      \hbox{\vrule width.#2pt height#1pt \kern#1pt
          \vrule width.#2pt}
      \hrule height.#2pt}}}
\def\bra0{\langle 0|}
\def\ket0{|0\rangle}

\definecolor{dred}{rgb}{0.7,0.15,0.09}
\definecolor{dblue}{rgb}{0,0.12,0.64}
\definecolor{dgreen}{rgb}{0.2,0.51,0.19}
\definecolor{pegn}{rgb}{0.33,0.51,0.14}

\newcommand{\pmat}[1]{\begin{pmatrix}#1\end{pmatrix}}

\newcommand{\mc}{\mathcal}
\newcommand{\mr}{\mathrm}

\newcommand{\del}{\partial}
\newcommand{\ol}{\overline}

\def\tr{\mathop{\mathrm{tr}}\nolimits}
\def\diag{\mathop{\mathrm{diag}}}

\newcommand{\sPhi}{\mathsf{\Phi}}

\newcommand{\utxtarrow}[1]{\xrightarrow[]{\hspace{1mm}#1\hspace{1mm}}}

\makeatletter
 
 \@addtoreset{equation}{section}
\makeatother

\begin{document}

\begin{titlepage}

\begin{flushright}
 KUNS-2865\\
 KYUSHU-HET-225\\
 KANAZAWA-21-04\\
\end{flushright}

\begin{center}

\vspace{1cm}
{\Large\textbf{
 Pseudo-Nambu-Goldstone Dark Matter Model\\
 Inspired by Grand Unification
}
 }
\vspace{1cm}

\renewcommand{\thefootnote}{\fnsymbol{footnote}}
Yoshihiko Abe$^{1}$\footnote[1]{y.abe@gauge.scphys.kyoto-u.ac.jp},
Takashi Toma$^{2,3}$\footnote[2]{toma@staff.kanazawa-u.ac.jp},
Koji Tsumura$^{4}$\footnote[3]{tsumura.koji@phys.kyushu-u.ac.jp},
Naoki Yamatsu${}^{4}$\footnote[4]{yamatsu.naoki@phys.kyushu-u.ac.jp}
\vspace{5mm}

\textit{
$^1${Department of Physics, Kyoto University, Kyoto 606-8502, Japan}\\
 $^2${Institute of Liberal Arts and Science,\\
 Kanazawa University, Kakuma-machi, Kanazawa, 920-1192 Japan}\\
 $^3$ Institute for Theoretical Physics, Kanazawa University, Kanazawa 920-1192, Japan
 $^4${Department of Physics, Kyushu University,\\
 744 Motooka, Nishi-ku, Fukuoka, 819-0395, Japan}
}

\vspace{8mm}

\abstract{
 A pseudo-Nambu-Goldstone boson (pNGB) is an attractive candidate for
 dark matter (DM) due to the simple evasion of the  current severe
 limits of DM direct detection experiments. 
 One of the pNGB DM models has been proposed based on a {\it gauged}
 $U(1)_{B-L}$ symmetry. The pNGB has long enough lifetime to be a DM
 and thermal relic abundance of pNGB DM can be fit with the observed
 value against the constraints on the DM decays from the cosmic-ray
 observations.
 The pNGB DM model can be embedded into an $SO(10)$ pNGB
 DM model in the framework of an $SO(10)$ grand unified theory,
 whose $SO(10)$ is broken to the Pati-Salam gauge group
 at the unified scale,
 and further to the Standard Model gauge group at the
 intermediate scale.
 Unlike the previous pNGB DM model, the parameters such as
 the gauge coupling constants of $U(1)_{B-L}$,
 the kinetic mixing parameter of between $U(1)_Y$ and $U(1)_{B-L}$
 are determined by solving the renormalization group equations
 for gauge coupling constants with  appropriate matching conditions.
 From the constraints of the DM lifetime and 
 gamma-ray observations, 
 the pNGB DM mass must be less than $\mathcal{O}(100)$$\,$GeV.
 We find that the thermal relic abundance can be consistent with all the
 constraints when the DM mass is close to half of the CP even Higg
 masses.
 }

\end{center}
\end{titlepage}

\section{Introduction}

The existence of dark matter (DM) has been confirmed by several
astronomical observations such as
spiral galaxies \cite{Corbelli:1999af,Sofue:2000jx},
gravitational lensing \cite{Massey:2010hh}, 
cosmic microwave background \cite{Aghanim:2018eyx},
and collision of bullet cluster \cite{Randall:2007ph}.
There are no viable DM candidates in the Standard Model (SM), so the
identification of DM plays an important role in particle physics as well
as cosmology.

Due to the lack of understanding the nature of DM, there are a lot of DM
candidates. One of the candidates is so-called Weakly Interacting
Massive Particle (WIMP). 
To realize the the relic abundance of DM, the WIMP mass is expected to
the range of $\mathcal{O}(10)$$\,$GeV to $\mathcal{O}(100)$$\,$TeV.
Further, since the WIMPs have non-gravitational interaction, the direct
and indirect detections are expected, but there are still no clear
signals of WIMPs, which lead to the strong constraint for WIMP mass and
interactions, especially from the direct detection.

Several mechanisms in WIMP DM models are proposed to avoid the severe
constrains from the direct detection by considering e.g.,
a fermion DM with pseudo-scalar interactions
\cite{Freytsis:2010ne,Ipek:2014gua,Arcadi:2017wqi,Sanderson:2018lmj,Abe:2018emu,Abe:2019wjw}
and a pseudo-Nambu-Goldstone boson (pNGB) DM
\cite{Barger:2010yn,Gross:2017dan,Ishiwata:2018sdi,Huitu:2018gbc,Cline:2019okt,Jiang:2019soj,Arina:2019tib,Abe:2020iph,Okada:2020zxo,Zhang:2021alu}.
Usually, in pNGB DM models, additional global $U(1)$ symmetry is assumed 
in an ad hoc manner.

In Refs.~\cite{Abe:2020iph,Okada:2020zxo}, a pNGB DM model is proposed based on 
$G_{\rm SM}\times U(1)_{B-L}$ gauge groups, where
$G_{\rm SM}:=SU(3)_C\times SU(2)_L\times U(1)_Y$.
Two complex scalars with $Q_{B-L}=+1$ and $+2$, denoted as $S$ and $\Phi$,
and three right-handed neutrinos due to the gauge anomaly cancellation 
are introduced.
The gauge symmetry is spontaneously broken via the nonvanishing
vacuum expectation value (VEV) of the scalar fields
$S$ and $\Phi$ as below:
\begin{align}
 G_{\rm SM}\times U(1)_{B-L}
 \ {\longrightarrow}\
 G_{\rm SM}.
\end{align}
The results in the model are summarized below.
The DM direct detection cross section is
naturally suppressed as the same as other pNGB DM models.
The pNGB can decay through the new high scale suppressed operators, but
the pNGB has a lifetime long enough to be a DM in the wide range
of the parameter space of the model.
The thermal relic abundance of pNGB DM can be fit with the 
observed value against the constraints on the DM decays from
the cosmic-ray observations. 

From other viewpoints, the charge quantization of $U(1)_Y$, the gauge
anomaly cancellation of $G_{\rm SM}$, and the almost SM gauge coupling
unification even in non-supersymmetric SM seem 
to imply the existence of grand unification \cite{Georgi:1974sy}.
The unification scale is expected to be $\mathcal{O}(10^{15}-10^{18})$$\,$GeV,
where the lower bound comes from the current non-observation of the
nucleon decay \cite{Heeck:2019kgr}
and the upper bound comes from the Planck scale.
Also, the tiny neutrino masses from the neutrino oscillation data seem
to suggest an intermediate scale $\mathcal{O}(10^{10}-10^{14})$$\,$GeV
through a see-saw mechanism \cite{Minkowski:1977sc}.

In this paper, we propose an $SO(10)$ pNGB DM model in the framework of 
grand unified theories (GUTs). Each Weyl fermion in
${\bf 16}$ of $SO(10)$ contains one generation of quarks and leptons,
which includes a right-handed neutrino \cite{Fritzsch:1974nn}. 
The SM Higgs and two complex scalar fields $S$ and $\Phi$ in
Refs.~\cite{Abe:2020iph,Okada:2020zxo} are assigned to a scalar field in ${\bf 10}$,
${\bf 16}$, and ${\bf \overline{126}}$ of $SO(10)$, respectively.
There are several symmetry breaking patterns of $SO(10)$ to 
$G_{\rm SM}\times U(1)_{B-L}$ as below.
\begin{align}
 SO(10)\ {\longrightarrow}\ 
 G_{I}\ {\longrightarrow}\
 G_{\rm SM}\times U(1)_{B-L}.
\label{Eq:Symmetry-breaking-pattern}
\end{align}
where $G_{I}$ stands for the intermediate gauge group such as
the Pati-Salam gauge group
$G_{\rm PS}:=SU(4)_C\times SU(2)_L\times SU(2)_R$
\cite{Pati:1974yy}
and a left-right gauge group
$G_{\rm LR}:=SU(3)_C\times SU(2)_L\times SU(2)_R\times U(1)_{B-L}$
\cite{Pati:1975ca,Mohapatra:1978fy}.
We mainly focus on the case of $G_I=G_{\rm PS}$, but
we also consider the possibility for such as $G_I=G_{\rm LR}$,
where the cases are not favored for a pNGB DM model under our
assumption and experimental constraints. 
(For more information about GUT model building in general, see, e.g.,
Refs.~\cite{Slansky:1981yr,Yamatsu:2015gut}.)

We discuss the following three things. First, the value of the gauge
kinetic mixing between $U(1)_Y$ and $U(1)_{B-L}$ is a free parameter
in e.g., the non-GUT pNGB DM models \cite{Abe:2020iph,Okada:2020zxo},
while that is determined mainly by the GUT gauge group in $SO(10)$
models. Second, gauge coupling unification can be achieved due to the
contribution from the additional scalar fields that contain a DM
candidate. Then the intermediate scale $M_I$, the unification scale
$M_U$, and the gauge coupling constant of $U(1)_{B-L}$
are fixed by using the renormalization group equations (RGEs) for gauge
coupling constants.
Third, the mass of the pNGB in the $SO(10)$ pNGB DM model is limited
to be $\mathcal{O}(10-100)$$\,$GeV from experimental constraints.

The paper is organized as follows.  
In Sec.~\ref{Sec:Model}, we introduce the $SO(10)$ pNGB DM model.
In Sec.~\ref{Sec:RGE}, we find gauge coupling unification determines 
mass scales and gauge coupling constants of the model. 
In Sec.~\ref{Sec:Dark-matter}, the constraints from experiments
are discussed.  Section~\ref{Sec:Summary} is devoted to
summary and discussions.

\section{The model}
\label{Sec:Model}

The model consists of an $SO(10)$ gauge field $\mathsf{A}_\mu$, fermions
in {\bf 16} of $SO(10)$,
a real scalar field in ${\bf 210}$ 
of $SO(10)$,
and complex scalar fields in ${\bf 10}$, ${\bf 16}$ and ${\bf \overline{126}}$
of $SO(10)$.
The $SO(10)$ gauge field contains $G_{\rm SM}$ and $U(1)_{B-L}$ gauge
fields. 
Each fermion in {\bf 16} of $SO(10)$ corresponds to quarks and leptons.
Scalar fields in ${\bf 10}$, ${\bf 16}$, and ${\bf \overline{126}}$ 
of $SO(10)$ include
the Higgs $H$, $S$ and $\Phi$, respectively. 
A scalar field in
${\bf 210}$ of $SO(10)$ is responsible
for breaking the $SO(10)$ symmetry to $G_{\rm PS}$.
The matter content in the $SO(10)$ model is summarized in 
Table~\ref{Tab:Matter_content-GUT-SO10}.
\footnote{
In this paper, we introduced a scalar in ${\bf 10}$ of $SO(10)$ as a
complex scalar. To reproduce the observed mass spectra of quarks and
leptons, it is discussed in e.g., Ref.~\cite{Bajc:2005zf} that 
only the real scalar in ${\bf 10}$ of $SO(10)$ has some tensions.
}

\begin{table}[tbh]
\begin{center}
\begin{tabular}{|c|c||c||c|c|c|c|c|}\hline
 \rowcolor[gray]{0.8}
 &{$\mathsf{A}_\mu$}
 &{$\mathsf{\Psi}_{\bf 16}$}
 &$\mathsf{\Phi}_{\bf 10}$&$\mathsf{\Phi}_{\bf 16}$&$\mathsf{\Phi}_{\overline{\bf 126}}$
 &{$\mathsf{\Phi}_{\bf 210}$}
 \\\hline
 $SO(10)$ &{${\bf 45}$}
 &{${\bf 16}$}
 &${\bf 10}$&${\bf 16}$&${\overline{\bf 126}}$
 &{${\bf 210}$}\\\hline
 $SL(2,\mathbb{C})$ &$(1/2,1/2)$
 &$(1/2,0)$
 &$(0,0)$&$(0,0)$&$(0,0)$&$(0,0)$
 \\\hline
\end{tabular}
 \caption{\small The matter content in the $SO(10)$ model is shown.
 }
\label{Tab:Matter_content-GUT-SO10}
\end{center}  
\end{table}

The Lagrangian is given by
\begin{align}
 {\cal L}&=
 \sum_{\bf y={\bf 10,16,\overline{126}}}\left(\mathsf{D}_\mu\mathsf{\Phi}_{\bf y}\right)^\dag
 \left(\mathsf{D}^\mu\mathsf{\Phi}_{\bf y}\right)
 +\frac{1}{2}
 \left(\mathsf{D}_\mu\mathsf{\Phi}_{\bf 210}\right)^T
 \left(\mathsf{D}^\mu\mathsf{\Phi}_{\bf 210}\right)
 +\sum_{a=1}^3
 \overline{\mathsf{\Psi}_{\bf 16}^{(a)}}i
 \cancel{\mathsf{D}}\mathsf{\Psi}_{\bf 16}^{(a)}
 -\frac{1}{2}\mbox{tr}
 \left[\mathsf{F}_{\mu\nu}\mathsf{F}^{\mu\nu}\right]
 \nonumber\\
 &-\left(\sum_{\bf y=10,\overline{126}}\sum_{a,b}\mathsf{y}_{\bf y}^{(ab)}
 \mathsf{\Phi}_{\overline{\bf y}}
 \left(\mathsf{\Psi}_{\bf 16}^{(a)}\mathsf{\Psi}_{\bf 16}^{(b)}\right)_{\bf y}
 +\mbox{h.c.}\right)
 -V\left(\left\{\mathsf{\Phi}_{\bf x}\right\}\right),
\label{Eq:Lagrangian-SO10}
\end{align}
where
$\mathsf{D}_\mu:=\partial_\mu+i\mathsf{g}\mathsf{A}_\mu$,
$\mathsf{F}_{\mu\nu}:=
\partial_\mu \mathsf{A}_\nu-\partial_\nu \mathsf{A}_\mu
+i\mathsf{g}[\mathsf{A}_\mu,\mathsf{A}_\nu]$.
The scalar potential
$V\left(\left\{\mathsf{\Phi}_{\bf x}\right\}\right)$
contains quadratic, cubic, and quartic coupling terms,
where ${\bf x}={\bf 10,16,\overline{126},210}$.

\begin{table}[htb]
\begin{center}
\begin{tabular}{|c||c|c|c|c|c|c|}\hline
 \rowcolor[gray]{0.8}
 &\multicolumn{6}{|c|}{$\mathsf{\Psi}_{\bf 16}$}
 \\\hline
 $SO(10)$ &\multicolumn{6}{|c|}{${\bf 16}$}
 \\\hline\hline
 \rowcolor[gray]{0.8}
 &\multicolumn{2}{|c|}{$\psi_{\bf (4,2,1)}$}
 &\multicolumn{4}{|c|}{$\psi_{{\bf (\overline{4},1,2)}}$}
 \\\hline
 $G_{\rm PS}$  &\multicolumn{2}{|c|}{${\bf (4,2,1)}$}
 &\multicolumn{4}{|c|}{${{\bf (\overline{4},1,2)}}$}
 \\\hline\hline
 \rowcolor[gray]{0.8}
 & $Q_L$ & $L$ 
 & $u_{R}^{c}$ & $d_{R}^{c}$ & $e_{R}^{c}$ & $\nu_{R}^c$
 \\\hline\hline
 $SU(3)_{c}$
 & ${\bf 3}$ & ${\bf 1}$ 
 & $\bar{\bf 3}$ & $\bar{\bf 3}$ & ${\bf 1}$ & ${\bf 1}$ 
 \\ \hline
 $SU(2)_{L}$
 & ${\bf 2}$ & ${\bf 2}$
 & ${\bf 1}$ & ${\bf 1}$ & ${\bf 1}$ & ${\bf 1}$ 
 \\ \hline
 $U(1)_{Y}$
 & $+1/6$ & $-1/2$ 
 & $-2/3$ & $+1/3$ & $+1$ & $0$ 
 \\\hline
 $U(1)_{B-L}$
 & $+1/3$ & $-1$
 & $-1/3$ & $-1/3$ & $+1$ & $+1$ 
 \\\hline
\end{tabular}
 \caption{\small The content of fermions in the $SO(10)$ model is shown
 in the $G_{\rm PS}=SU(4)_C\times SU(2)_L\times SU(2)_R$ basis,
 where the fermions belong to $(1/2,0)$ under $SL(2,\mathbb{C})$.
 The $U(1)_{B-L}$ charge $Q_{B-L}$ is given by
 $U(1)(\subset SU(4)/SU(3)$) \cite{Yamatsu:2015gut}.
 }
\label{Tab:Matter_content-GUT-fermion}
\end{center}  
\end{table}

We consider the following symmetry breaking patterns of $SO(10)$ broken
to $G_{\rm PS}$ at the unification scale $M_U$ by
the nonvanishing vacuum expectation value (VEV) of the scalar field in
${\bf 210}$ in $SO(10)$, further to
$G_{\rm SM}$ at the intermediate scale $M_I$
by the VEV of the scalar field in ${\bf \overline{126}}$ in $SO(10)$,
where the $M_U$ and $M_I$ will be determined by gauge coupling
unification using the renormalization group equations (RGEs) for the
gauge coupling constants in the next section.
\begin{align}
 SO(10)
 &\utxtarrow{\langle\mathsf{\Phi}_{\bf 210}\rangle\not=0}
 G_{\rm PS}
 \left(
 \supset
 G_{\rm SM}\times U(1)_{B-L}\right)
 \utxtarrow{\langle\mathsf{\Phi}_{\bf \overline{126}}\rangle\not=0}
 G_{\rm SM}
 \utxtarrow{\langle\mathsf{\Phi}_{\bf 10}\rangle\not=0}
 SU(3)_C\times U(1)_{\rm EM},
 \label{Eq:SO10-Symmetry-breaking}
\end{align}
where the dominant contribution for the symmetry breaking from the
VEVs are shown.
The type of symmetry breaking has been already discussed in 
e.g.,
Refs.~\cite{Fritzsch:1974nn,Aulakh:1982sw,Babu:1992ia,Aulakh:2003kg,Bajc:2005zf,Fukuyama2005,Bertolini:2009qj,Altarelli:2013aqa,Fukuyama:2012rw,Mambrini:2015vna,Ellis:2018khn,Ferrari:2018rey,Chakrabortty:2019fov,Chakraborty:2019uxk}.
The field content of fermion, scalar, and gauge bosons are shown in 
Tables~\ref{Tab:Matter_content-GUT-fermion},  
\ref{Tab:Matter_content-GUT-scalar}, and
\ref{Tab:Matter_content-GUT-gauge}.
(The potential analysis of ${\bf 210}$ in $SO(10)$ has already discussed
in e.g., Ref.~\cite{Chang:1985zq}; $SO(10)$ is broken to $G_{\rm PS}$
for appropriate parameter sets.)

\begin{table}[htb]
\begin{center}
\begin{tabular}{|c||c|c|c|c|c|}\hline
 \rowcolor[gray]{0.8}
 &$\mathsf{\Phi}_{\bf 10}$&$\mathsf{\Phi}_{\bf 16}$&$\mathsf{\Phi}_{\overline{\bf 126}}$
 \\\hline
 $SO(10)$
 &${\bf 10}$ & ${\bf 16}$ & $\overline{\bf 126}$
 \\\hline\hline
 \rowcolor[gray]{0.8}
 &$\phi_{\bf (1,2,2)}$
 &$\phi_{\bf (\overline{4},1,2)}$
 &$\phi_{\bf (\overline{10},1,3)}$
 \\\hline
 $G_{\rm PS}$
 &${\bf (1,2,2)}$ & ${\bf (\overline{4},1,2)}$
 & ${\bf (\overline{10},1,3)}$
 \\\hline\hline
 \rowcolor[gray]{0.8}
 & $H$ & $S$ & $\Phi$
 \\\hline\hline
 $SU(3)_{c}$
 & ${\bf 1}$ & ${\bf 1}$ & ${\bf 1}$
 \\ \hline
 $SU(2)_{L}$
 & ${\bf 2}$ & ${\bf 1}$ & ${\bf 1}$
 \\ \hline
 $U(1)_{Y}$
 & $+1/2$ & $0$ & $0$
 \\\hline
 $U(1)_{B-L}$
 & $0$ & $+1$ & $+2$
 \\\hline
\end{tabular}
 \caption{\small The content of scalar fields in the $SO(10)$ model is
 shown,  where the scalars belong to $(0,0)$ under $SL(2,\mathbb{C})$;
 $\mathsf{\Phi}_{\bf 10}$,
 $\mathsf{\Phi}_{\bf 16}$ and $\mathsf{\Phi}_{\bf \overline{126}}$
 are complex scalar  fields.
 Here we assume all unlisted components of $G_{\rm PS}$ have
 ${\cal O}(M_{U})$ masses and also all unlisted components of
 $G_{\rm SM}\times U(1)_{B-L}$ have $\mathcal{O}(M_{I})$ and $\mathcal{O}(M_{U})$ 
 masses, respectively.
 Other information is the same as in Table~\ref{Tab:Matter_content-GUT-fermion}.
 }
\label{Tab:Matter_content-GUT-scalar}
\end{center}  
\end{table}

\begin{table}[htb]
\begin{center}
\begin{tabular}{|c||c|c|c|c|}\hline
 \rowcolor[gray]{0.8}
 &\multicolumn{4}{|c|}{$\mathsf{A}_\mu$}
 \\\hline
 $SO(10)$
 &\multicolumn{4}{|c|}{${\bf 45}$}
 \\\hline\hline
 \rowcolor[gray]{0.8}
 &\multicolumn{2}{|c|}{$G_\mu'$}&$W_\mu$&$W_\mu'$
 \\\hline
 $G_{\rm PS}$
 &\multicolumn{2}{|c|}{${\bf (15,1,1)}$}
 &${\bf (1,3,1)}$
 &${\bf (1,1,3)}$
 \\\hline\hline
 \rowcolor[gray]{0.8}
 & $G_\mu$ &$C_\mu$& $W_\mu$ & $Z_\mu'$
 \\\hline\hline
 $SU(3)_{c}$
 & ${\bf 8}$ & ${\bf 1}$ & ${\bf 1}$ & ${\bf 1}$
 \\ \hline
 $SU(2)_{L}$
 & ${\bf 1}$ & ${\bf 1}$ & ${\bf 3}$ & ${\bf 1}$
 \\ \hline
 $U(1)_{Y}$
 & $0$ & $0$ & $0$ & $0$
 \\\hline
 $U(1)_{B-L}$
 & $0$ & $0$ & $0$ & $0$
 \\\hline
\end{tabular}
 \caption{\small The content of gauge fields in the $SO(10)$ model is
 shown, where the gauge fields belong to $(1/2,1/2)$ under
 $SL(2,\mathbb{C})$; 
 Other information is the same as in
 Tables~\ref{Tab:Matter_content-GUT-fermion} and
 \ref{Tab:Matter_content-GUT-scalar}.
 }
\label{Tab:Matter_content-GUT-gauge}
\end{center}  
\end{table}

\subsection{Scalar sector}
\label{Sec:Scalar}

Here we focus on  the scalar potential of SM Higgs and pNGB relevant
part that contains scalar fields $H$, $S$, $\Phi$ belonging to  
${\bf 10}$, ${\bf 16}$, and ${\bf \overline{126}}$ of $SO(10)$,
respectively.
We assume that the other components of
$\mathsf{\Phi}_{\bf 10}$, $\mathsf{\Phi}_{\bf 16}$ and 
$\mathsf{\Phi}_{\overline{\bf 126}}$
shown in Table~\ref{Tab:Matter_content-GUT-scalar}
have the intermediate scale or larger masses and they do not contribute
$SU(2)_L\times U(1)_Y$ and $U(1)_{B-L}$ breakings.

From the scalar potential
$V\left(\left\{\mathsf{\Phi}_{\bf x}\right\}\right)$
in Eq.~(\ref{Eq:Lagrangian-SO10}), we extract the terms that contain
only $H$, $S$, $\Phi$:
\begin{align}
 V\left(H,S,\Phi\right)
 &=-\frac{\mu_H^2}{2}|H|^2
 -\frac{\mu_S^2}{2}|S|^2
 -\frac{\mu_\Phi^2}{2}|\Phi|^2
 +\frac{\lambda_H}{2}|H|^4
 +\frac{\lambda_S}{2}|S|^4
 +\frac{\lambda_\Phi}{2}|\Phi|^4
 \nonumber\\
 &+\lambda_{HS}|H|^2|S|^2
 +\lambda_{H\Phi}|H|^2|\Phi|^2
 +\lambda_{S\Phi}|S|^2|\Phi|^2
 -\left(
 \frac{\mu_c}{\sqrt{2}}\Phi^*S^2+\mbox{c.c.}
 \right).
 \label{Eq:Scalar-potential}
\end{align}
The quadratic terms $|H|^2$, $|S|^2$, and $|\Phi|^2$
come from
$(\mathsf{\Phi}_{\bf 10}\mathsf{\Phi}_{\bf 10})_{\bf 1}$,
$(\mathsf{\Phi}_{\bf 16}\mathsf{\Phi}_{\bf 16}^*)_{\bf 1}$, and
$(\mathsf{\Phi}_{\bf \overline{126}}\mathsf{\Phi}_{\bf
\overline{126}}^*)_{\bf 1}$, respectively;
the quartic terms $|H|^4$, $|S|^4$, and $|\Phi|^4$
come from
$((\mathsf{\Phi}_{\bf 10}\mathsf{\Phi}_{\bf 10})_{\bf 1})^2$
and 
$|(\mathsf{\Phi}_{\bf 10}\mathsf{\Phi}_{\bf 10})_{\bf 54}|^2$ 
$|(\mathsf{\Phi}_{\bf 16}\mathsf{\Phi}_{\bf 16})_{\bf \overline{126}}|^2$,
and 
$|(\mathsf{\Phi}_{\bf \overline{126}}\mathsf{\Phi}_{\bf \overline{126}})_{\bf \overline{2772}}|^2$, respectively;
the quartic terms $|H|^2|S|^2$, $|H|^2|\Phi|^2$, and $|S|^2|\Phi|^2$
come from
$(\mathsf{\Phi}_{\bf 10}\mathsf{\Phi}_{\bf 10})_{\bf 1}
(\mathsf{\Phi}_{\bf 16}\mathsf{\Phi}_{\bf 16}^*)_{\bf 1}$, 
$(\mathsf{\Phi}_{\bf 10}\mathsf{\Phi}_{\bf 10})_{\bf 1}
(\mathsf{\Phi}_{\bf \overline{126}}\mathsf{\Phi}_{\bf \overline{126}}^*)_{\bf 1}$,
and
$(\mathsf{\Phi}_{\bf 16}\mathsf{\Phi}_{\bf 16}^*)_{\bf 1}
(\mathsf{\Phi}_{\bf \overline{126}}\mathsf{\Phi}_{\bf \overline{126}}^*)_{\bf 1}$,
respectively;
the cubic term $\Phi^*S^2$
comes from 
$\mathsf{\Phi}_{\bf \overline{126}}^*
(\mathsf{\Phi}_{\bf 16}\mathsf{\Phi}_{\bf 16})_{\bf \overline{126}}$,
\footnote{
When we take into account the nonvanishing VEV of
$\mathsf{\Phi}_{\bf 210}$,
quadratic terms $|H|^2$, $|S|^2$, and $|\Phi|^2$ and 
the cubic term $\Phi^*S^2$ also come from
$(\mathsf{\Phi}_{\bf 10}\mathsf{\Phi}_{\bf 10})_{\bf 1}
(\mathsf{\Phi}_{\bf 210}\mathsf{\Phi}_{\bf 210})_{\bf 1}$,
$(\mathsf{\Phi}_{\bf 16}\mathsf{\Phi}_{\bf 16}^*)_{\bf 1}
(\mathsf{\Phi}_{\bf 210}\mathsf{\Phi}_{\bf 210})_{\bf 1}$, 
$(\mathsf{\Phi}_{\bf \overline{126}}\mathsf{\Phi}_{\bf
\overline{126}}^*)_{\bf 1}
(\mathsf{\Phi}_{\bf 210}\mathsf{\Phi}_{\bf 210})_{\bf 1}$, 
$\mathsf{\Phi}_{\bf 16}\mathsf{\Phi}_{\bf 16}\mathsf{\Phi}_{\bf \overline{126}}\mathsf{\Phi}_{\bf 210}$,
respectively.
Therefore, each coefficient such as $\mu_c$ in Eq.~(\ref{Eq:Scalar-potential}) should be
regarded as the total value including all the corresponding terms such
as
$\mathsf{\Phi}_{\bf 16}\mathsf{\Phi}_{\bf 16}\mathsf{\Phi}_{\bf \overline{126}}$
and $\mathsf{\Phi}_{\bf 16}\mathsf{\Phi}_{\bf 16}\mathsf{\Phi}_{\bf \overline{126}}\mathsf{\Phi}_{\bf 210}$.
}
where the above subscript such as ${\bf 1}$ and ${\bf 54}$ stands for
the product representation of $SO(10)$.
This potential is exactly the same as that in Refs.~\cite{Abe:2020iph,Okada:2020zxo}.

We assume that the scalar fields $H$, $S$, and $\Phi$ develop the VEVs, 
which are parameterized by
\begin{align}
 H=
 \left(
\begin{array}{c}
 0\\
 \frac{v+h}{\sqrt{2}}\\
\end{array}
 \right),\ \ \
 S=\frac{v_s+s+i\eta_s}{\sqrt{2}},\ \ \
 \Phi=\frac{v_\phi+\phi+i\eta_\phi}{\sqrt{2}},
\label{Eq:Scalar-VEVs} 
\end{align}
where $h$, $s$, and $\phi$ are CP-even modes, 
$\eta_s$ and $\eta_\phi$ are CP-odd modes,
and $v$, $v_s$, and $v_\phi$ are the VEVs of $H$, $S$, and $\Phi$,
respectively.
The CP phase of the cubic term $\Phi^*S^2$ is eliminated by the field
redefinition of $\Phi$. In the limit $\mu_c\to 0$, there are two
independent global $U(1)$ symmetries associated with the phase rotation
of $S$ and $\Phi$. For $\mu_c\not=0$, the $U(1)$ symmetries are merged
to the $U(1)_{B-L}$ (or $U(1)_X$) symmetry. Once $U(1)_{B-L}$ is broken,
one of two CP-odd modes is absorbed by the $U(1)_{B-L}$ gauge field denoted
as $C_\mu$, while the other appears as a physical pNGB whose mass is
proportional to $\mu_c$.

The scalar fields $H$, $S$, $\Phi$ have five modes; three of
them are CP-even scalar modes and the other two are CP-odd modes.
The mass matrix for the CP-even scalars in the $(h,s,\phi)$ basis is
given by
\begin{align}
 M_{\rm even}^2=
 \left(
 \begin{array}{ccc}
  \lambda_H v^2&\lambda_{HS}vv_s&\lambda_{H\Phi}vv_\phi \\
  \lambda_{HS}vv_s&\lambda_S v_s^2&\lambda_{S\Phi}v_sv_\phi-\mu_cv_s\\
  \lambda_{H\Phi}vv_\phi&\lambda_{S\Phi}v_sv_\phi-\mu_cv_s
   &\lambda_\Phi v_\phi^2+\frac{\mu_cv_s^2}{2v_\phi}\\
 \end{array}
 \right).
\end{align}
Since the matrix is real and symmetric, it can be diagonalized by
a real orthogonal matrix.
The gauge eigenstates $(h,s,\phi)$ are related with the mass eigenstates
$(h_1,h_2,h_3)$ as
\begin{align}
 \left(
 \begin{array}{c}
  h\\
  s\\
  \phi\\
 \end{array}
 \right)=
 U_e
 \left(
 \begin{array}{c}
  h_1\\
  h_2\\
  h_3\\
 \end{array}
 \right),
\end{align}
where the approximate form of the real orthogonal matrix and its mixing
angle are given by
\begin{align}
 &U_e\simeq
 \left(
 \begin{array}{ccc}
  1&0&\frac{\lambda_{H\Phi}v}{\lambda_\Phi v_\phi}\\
  0&1&\frac{\lambda_{S\Phi}v}{\lambda_\Phi v_\phi}\\
  -\frac{\lambda_{H\Phi}v}{\lambda_\Phi v_\phi}
   &-\frac{\lambda_{S\Phi}v}{\lambda_\Phi v_\phi}&1\\ 
 \end{array}
 \right)
 \left(
 \begin{array}{ccc}
  \cos\theta&\sin\theta&0\\
  -\sin\theta&\cos\theta&0\\
  0&0&1\\
 \end{array}
 \right),\\
 &\tan2\theta
 \simeq
 \frac{2vv_s(\lambda_{HS}\lambda_\Phi-\lambda_{H\Phi}\lambda_{S\Phi})}
 {v^2(\lambda_{H\Phi}^2-\lambda_H\lambda_\Phi)
 -v_s^2(\lambda_{S\Phi}^2-\lambda_S\lambda_\Phi)}.
 \label{Eq:U-even}
\end{align}
The masses of $(h_1,h_2,h_3)$ are given by
\begin{align}
 m_{h_1}^2&\simeq
 \lambda_Hv^2-\frac{\lambda_{H\Phi}^2\lambda_S
 -2\lambda_{HS}\lambda_{H\Phi}\lambda_{S\Phi}+\lambda_\Phi\lambda_{HS}^2}
 {\lambda_S\lambda_\Phi-\lambda_{S\Phi}^2}v^2,\\
 m_{h_2}^2&\simeq
 \frac{\lambda_S\lambda_\Phi-\lambda_{S\Phi}^2}{\lambda_\Phi}v_s^2
 +\frac{(\lambda_\Phi\lambda_{HS}-\lambda_{H\Phi}\lambda_{S\Phi})^2}
 {\lambda_\Phi(\lambda_S\lambda_\Phi-\lambda_{S\Phi}^2)}v^2,\\
 m_{h_3}^2&\simeq
 \lambda_{\Phi}v_\phi^2.
\end{align}
The mass eigenstate $h_1$ is identified as the SM-like Higgs boson with
the mass 
$m_{h_1}\simeq 125$$\,$GeV,
$h_2$ is a light CP-even scalar, and
$h_3$ is a heavy CP-even scalar.

The mass matrix of the CP-odd scalars in the gauge eigenstates
$(\eta_s,\eta_\phi)$ is given by
\begin{align}
 M_{\rm odd}^2=\frac{\mu_c}{2v_\phi}
 \left(
 \begin{array}{cc}
  4v_\phi^2&-2v_sv_\phi\\
  -2v_sv_\phi&v_s^2\\
 \end{array}
 \right).
\end{align}
The gauge eigenstates $(\eta_s,\eta_\phi)$ are related
with the mass eigenstates $(\chi,\tilde{\chi})$ as
\begin{align}
 \left(
 \begin{array}{c}
  \eta_s\\
  \eta_\phi\\
 \end{array}
 \right)=
 U_o
 \left(
 \begin{array}{c}
  \chi\\
  \tilde{\chi}\\
 \end{array}
 \right),
\end{align}
where the real orthogonal matrix is given by
\begin{align}
 &U_o=\frac{1}{\sqrt{v_s^2+4v_\phi^2}}
 \left(
 \begin{array}{cc}
  2v_\phi&v_s\\
  -v_s&2v_\phi\\
 \end{array}
 \right).
 \label{Eq:U-odd}
\end{align}
By using the $2\times 2$ real orthogonal matrix $U_o$, the mass
eigenvalues of $(\chi,\tilde{\chi})$ are given by
\begin{align}
m_\chi^2&=\frac{(v_s^2+4v_\phi^2)\mu_c}{4v_\phi},\\
m_{\tilde{\chi}}^2&=0.
\end{align}
The $\tilde{\chi}$ is the NGB absorbed by the $U(1)_{B-L}$ gauge boson
$C_\mu$, and $\chi$ is the pNGB identified as DM in the paper.

\subsection{Gauge sector}

The gauge kinetic term of the $SO(10)$ can be canonically normalized at
the unification scale $M_U$ as in  Eq.~(\ref{Eq:Lagrangian-SO10}).
In general, the kinetic-mixing term of multiple $U(1)$ symmetries are allowed for
the case of at least two abelian groups because a field strength itself is
gauge-invariant for abelian groups, while that is not gauge-invariant
for non-abelian groups. 
So, in the energy scale $M_I<\mu<M_U$, there is the gauge kinetic mixing
of $G_{\rm PS}$.
At the scale $\mu=M_{I}$, there are two $U(1)$s, i.e. $U(1)_Y$  and
$U(1)_{B-L}$ although one of the $U(1)$s,
which is the $U(1)_{B-L}$, is broken at the scale.
It is generated by threshold corrections or via RGE flows.
In $SO(10)$ models, $SO(10)/(SU(3)_C\times SU(2)_L)$ contains $U(1)_Y$
and $U(1)_{B-L}$ as two independent $U(1)$s, while they are not
orthogonal.
In fact, $U(1)_Y$ is orthogonal to $U(1)_X(\subset SO(10)/SU(5))$;
$U(1)_{B-L}$ is orthogonal to $U(1)_R(\subset SU(2)_R$).
Therefore, it is expected that the kinetic mixing parameter between
$U(1)_Y$ and $U(1)_{B-L}$ denoted as $\epsilon$ is non-zero at
classical level.

To determine the value of the kinetic mixing parameter between $U(1)_Y$
and $U(1)_{B-L}$, we focus on the kinetic terms of the gauge
fields.
First, from Eq.~(\ref{Eq:Lagrangian-SO10}), the gauge kinetic term of
$SO(10)$ is given by
\begin{align}
 {\cal L}_{\rm gauge}=
 -\frac{1}{2}\mbox{tr}
 \left[\mathsf{F}_{\mu\nu}\mathsf{F}^{\mu\nu}\right].
\label{Eq:Lagrangian-SO10-GK} 
\end{align}
Next, the gauge kinetic terms of $G_{\rm PS}$ are given by
\begin{align}
 {\cal L}_{\rm gauge}
 \ni
 -\frac{1}{2}\mbox{tr}\left[{G}_{\mu\nu}^{\prime}{G}^{\prime\mu\nu}\right]
 -\frac{1}{4}{W}_{\mu\nu}^a{W}^{a\mu\nu}
 -\frac{1}{4}{W}_{\mu\nu}^{\prime a}{W}^{\prime a\mu\nu},
\label{Eq:Lagrangian-SO10-GPS} 
\end{align}
where ${G}_{\mu\nu}^{\prime}$, ${W}_{\mu\nu}^a$, and
${W}_{\mu\nu}^{\prime a}$ stand for
the field strengths of $SU(4)_C$, $SU(2)_L$, and $SU(2)_R$, respectively;
the gauge kinetic terms and mass terms of $SO(10)/G_{\rm PS}$ are
omitted at $M_U$.
The gauge coupling constants are running from $M_U$ to $M_I$.
Third, the $SU(3)_C\times SU(2)_L\times U(1)_R\times U(1)_{B-L}$ are
given by
\begin{align}
 {\cal L}_{\rm gauge}
 \ni
 -\frac{1}{2}\mbox{tr}\left[{G}_{\mu\nu}{G}^{\mu\nu}\right]
 -\frac{1}{4}{W}_{\mu\nu}^a{W}^{a\mu\nu}
 -\frac{1}{4}{B}_{\mu\nu}^{\prime}{B}^{\prime\mu\nu}
 -\frac{1}{4}{C}_{\mu\nu}^{\prime}{C}^{\prime\mu\nu},
\label{Eq:Lagrangian-SO10-G3211-1}
\end{align}
where ${G}_{\mu\nu}$, ${B}_{\mu\nu}^{\prime}$ and ${C}_{\mu\nu}^{\prime}$
stand for the field strength of $SU(3)_C(\subset SU(4)_C)$,
$U(1)_R(\subset SU(2)_R)$, and $U(1)_{B-L}(\subset SU(4)_C/SU(3)_C)$,
respectively;
the gauge kinetic terms and mass terms of
$SU(4)_C/(SU(3)_C\times U(1)_{B-L})$ and $SU(2)_R/U(1)_R$ are
omitted at $M_I$.
Further, by using the following $GL(2,\mathbb{R})$ transformation
\begin{align}
 \begin{array}{l}
 U(1)_Y\\
 U(1)_{B-L}\\
 \end{array}
 :\ &
 \left(
 \begin{array}{c}
 B_\mu\\
 C_\mu\\
 \end{array}
 \right)=
 \left(
 \begin{array}{cc}
 1&-\tan\epsilon\\
 0&\frac{1}{\cos\epsilon}\\
 \end{array}
 \right)
 \left(
 \begin{array}{cc}
 B_\mu'\\
 C_\mu'\\
 \end{array}
 \right)=:U_{GK}
 \left(
 \begin{array}{cc}
 B_\mu'\\
 C_\mu'\\
 \end{array}
 \right)\ \ \ :
 \begin{array}{l}
 U(1)_R\\
 U(1)_{B-L}\\
 \end{array},
 \label{Eq:U1Y-BL_R-BL}
\end{align}
we can change the basis of $U(1)$s from
$U(1)_{R}\times U(1)_{B-L}$ to $U(1)_{Y}\times U(1)_{B-L}$;
\begin{align}
 -\frac{1}{4}{B}_{\mu\nu}^{\prime}{B}^{\prime\mu\nu}
 -\frac{1}{4}{C}_{\mu\nu}^{\prime}{C}^{\prime\mu\nu}
 =-\frac{1}{4}{B}_{\mu\nu}{B}^{\mu\nu}
 -\frac{1}{4}{C}_{\mu\nu}{C}^{\mu\nu}
 -\frac{\sin\epsilon}{2}{C}_{\mu\nu}{B}^{\mu\nu},
\label{Eq:Lagrangian-SO10-G3211-2}
\end{align}
where ${B}_{\mu\nu}$ and $C_{\mu\nu}$ stand for the field strength of
$U(1)_Y$ and $U(1)_{B-L}$, respectively;
$\epsilon$ is the kinetic mixing parameter between $U(1)_Y$ and
$U(1)_{B-L}$. In the case, since the $U(1)_Y$ generator is given 
by the following linear combination of $U(1)_R$ and $U(1)_{B-L}$
\begin{align}
 I_Y=\sqrt{\frac{3}{5}}I_{3R}+\sqrt{\frac{2}{5}}I_{B-L}.
 \label{Eq:U1_Y}
\end{align}
Due to the orthogonality, the kinetic mixing parameter $\epsilon$ at
$\mu=M_I$ is given by 
\begin{align}
\epsilon=-\tan^{-1}\sqrt{\frac{2}{3}}.
\label{Eq:kinetic-mixing-parameter}
\end{align}

The Lagrangian for the electro-magnetic neutral part of
the $SU(2)_L\times U(1)_Y\times U(1)_{B-L}$
gauge fields including mass terms generated by the VEVs of the
spontaneous $SU(2)_L\times U(1)_Y$ and $U(1)_{B-L}$ breaking scalar
fields is given by 
\begin{align}
 {\cal L}&=
 -\frac{1}{4}{B}_{\mu\nu}{B}^{\mu\nu}
 -\frac{1}{4}{W}_{\mu\nu}^3{W}^{3\mu\nu}
 +\frac{1}{2}M_{\bar{Z}}^2{Z}_{\mu}{Z}^{\mu}
 \nonumber\\
 &
 -\frac{1}{4}{C}_{\mu\nu}{C}^{\mu\nu}
 +\frac{1}{2}M_C^2{C}_{\mu}{C}^{\mu}
 -\frac{\sin\epsilon}{2}{C}_{\mu\nu}{B}^{\mu\nu},
\label{Eq:Lagrangian-gause-kinetic-mass}
\end{align}
where 
$Z_\mu=\cos\theta_W W_\mu^3-\sin\theta_WB_\mu$ is the usual $Z$ boson,
$\theta_W$ is the Weinberg angle $\tan\theta_W:=g_{1}/g_{2}$;
$g_{1}$ and $g_2$ stand for the
$U(1)_Y$ and $SU(2)_L$ coupling constants, respectively.
The mass parameters are given by
\begin{align}
 &M_{\bar{Z}}^2=\frac{g_{1}^2+g_{2}^{2}}{4}v^2,\ \ \
 M_C^2=g_{B-L}^2(v_s^2+4v_\phi^2),
\end{align}
where $g_{B-L}$ is the gauge coupling constant of $U(1)_{B-L}$.

To discuss the physical implications of $U(1)_{B-L}$ gauge boson, we 
requires both diagonalizing the field strength terms and the mass
terms.
First, we diagonalize the kinetic term in
Eq.~(\ref{Eq:Lagrangian-gause-kinetic-mass}) by using the following
$GL(2,\mathbb{R})$ transformation:
\begin{align}
 \begin{array}{l}
 U(1)_Y\\
 U(1)_{B-L}\\
 \end{array}
 :\ &
 \left(
 \begin{array}{c}
 B_\mu\\
 C_\mu\\
 \end{array}
 \right)=
 \left(
 \begin{array}{cc}
 1&-\tan\epsilon\\
 0&\frac{1}{\cos\epsilon}\\
 \end{array}
 \right)
 \left(
 \begin{array}{cc}
 \hat{B}_\mu\\
 \hat{C}_\mu\\
 \end{array}
 \right)=U_{GK}
 \left(
 \begin{array}{cc}
 \hat{B}_\mu\\
 \hat{C}_\mu\\
 \end{array}
 \right),
 \label{Eq:GL2R-transformation}
\end{align}
where $\hat{B}_\mu$ and $\hat{C}_\mu$ stand for
the gauge fields of the $U(1)_Y$ and ``$U(1)_{B-L}$'' in the physical
basis. The transformation is exactly the same as that in
Eq.~(\ref{Eq:U1Y-BL_R-BL}).
That is, ``$U(1)_{B-L}$'' can be identified as
$U(1)_X(\subset SO(10)/SU(5))$.
Then, the gauge kinetic terms in
Eq.~(\ref{Eq:Lagrangian-gause-kinetic-mass}) 
become 
\begin{align}
 {\cal L}_{\rm GK}&=
 -\frac{1}{4}{\hat{B}}_{\mu\nu}{\hat{B}}^{\mu\nu}
 -\frac{1}{4}{\hat{W}}_{\mu\nu}^3{\hat{W}}^{3\mu\nu}
 -\frac{1}{4}{\hat{C}}_{\mu\nu}{\hat{C}}^{\mu\nu}.
\label{Eq:Lagrangian-gause-kinetic-mass-no-kinetic-mixing}
\end{align}
Next, we consider the physical eigenstate via an $O(3)$ rotation by
diagonalizing the mass terms that arise after both $U(1)_{B-L}$ 
and $SU(2)_L\times U(1)_Y$ breaking. One mass eigenstate is massless
corresponding to the photon $A_\mu$, while the other two denoted
$Z$ and $Z'$ receive masses.
The mass terms of the neutral gauge boson in terms of
$(B_\mu,W_\mu^3,C_\mu)$ is given by
\begin{align}
 {\cal L}_{\rm mass}
 =\frac{1}{2}
 \left(B_\mu\ W_\mu^3\ C_\mu\right)
 \left(
 \begin{array}{ccc}
  \sin^2\theta_W M_{\bar{Z}}^2&-\sin\theta_W\cos\theta_W M_{\bar{Z}}^2&0\\
  -\sin\theta_W \cos\theta_W M_{\bar{Z}}^2&\cos^2\theta_WM_{\bar{Z}}^2&0\\
  0&0&M_C^2\\
 \end{array}
 \right)
 \left(
 \begin{array}{c}
  B^\mu\\
  W^{3\mu}\\
  C^\mu\\
 \end{array}
 \right).
\end{align}
By using $GL(2,\mathbb{R})$ transformation in
Eq.~(\ref{Eq:GL2R-transformation}), we change the basis whose kinetic
term is diagonalized as below:
\begin{align}
 {\cal L}_{\rm mass}
 &=\frac{1}{2}
 \left(\hat{B}_\mu\ W_\mu^3\ \hat{C}_\mu\right)
 \tilde{U}_{GK}^T
 \left(
 \begin{array}{ccc}
  \sin^2\theta_W M_{\bar{Z}}^2&-\sin\theta_W\cos\theta_WM_{\bar{Z}}^2&0\\
  -\sin\theta_W\cos\theta_WM_{\bar{Z}}^2&\cos^2\theta_WM_{\bar{Z}}^2&0\\
  0&0&M_C^2\\
 \end{array}
 \right)
 \tilde{U}_{GK}
 \left(
 \begin{array}{c}
  \hat{B}^\mu\\
  W^{3\mu}\\
  \hat{C}^\mu\\
 \end{array}
 \right),
\end{align}
where
\begin{align}
 \tilde{U}_{GK}:=
 \left(
 \begin{array}{ccc}
 1&0&-\tan\epsilon\\
 0&1&0\\
 0&0&\frac{1}{\cos\epsilon}\\
 \end{array}
 \right).
\end{align}
The above mass matrix is a real symmetric matrix. In fact,
it can be diagonalized by using a real orthogonal matrix:
\begin{align}
 U_G=
  \left(
 \begin{array}{ccc}
 \cos\theta_W&-\sin\theta_W&0\\
 \sin\theta_W&\cos\theta_W&0\\
 0&0&1\\
 \end{array}
 \right)
  \left(
 \begin{array}{ccc}
 1&0&0\\
 0&\cos\zeta&-\sin\zeta\\
 0&\sin\zeta&\cos\zeta\\
 \end{array}
 \right),
\end{align}
where the mixing angle $\zeta$ is given by
\begin{align}
 \tan2\zeta=
 \frac{-2M_Z^2\sin\theta_W \sin\epsilon \cos\epsilon}
 {M_C^2-M_Z^2(\cos^2\epsilon-\sin^2\theta_W\sin^2\epsilon)}.
\end{align}
From the above, we find the masses of $A_\mu$, $Z_\mu$, and $Z_\mu'$ as 
\begin{align}
 M_A^2&=0,\\
 M_Z^2&=\frac{1}{2}
 \left[
 \overline{M}^2-\sqrt{\overline{M}^4-\frac{4M_{\bar{Z}}^2M_C^2}{\cos^2\epsilon}}
 \right],\\
 M_{Z'}^{2}&=\frac{1}{2}
 \left[
 \overline{M}^2+\sqrt{\overline{M}^4-\frac{4M_{\bar{Z}}^2M_C^2}{\cos^2\epsilon}}
 \right],
\end{align}
where $\overline{M}^2$ is given by
\begin{align}
\overline{M}^2:=M_{\bar{Z}}^2\left(1+\sin\theta_W\tan^2\epsilon\right)+\frac{M_C^2}{\cos^2\epsilon}.
\end{align}

In this section, we find that the gauge kinetic mixing $\epsilon$ in
Refs.~\cite{Abe:2020iph,Okada:2020zxo} is regarded as the mixing angle.
In Appendix~\ref{Sec:Kinetic-mixing}, we will show this more explicitly.

\section{Gauge coupling constants}
\label{Sec:RGE}

To determine such as the $U(1)_{B-L}$ breaking scale, i.e., intermediate
scale $M_I$, and magnitude of the gauge coupling constant of the
$U(1)_{B-L}$, we 
discuss the RGEs for gauge coupling constants running among  the
electroweak scale $M_Z$, the intermediate scale $M_{I}$, and the
unification scale $M_U$.

The RGE for the gauge coupling constant
$\alpha_i(\mu):= g_i^2(\mu)/4\pi$ at one-loop level is given in e.g.,
Refs.~\cite{Slansky:1981yr,Yamatsu:2015gut} by 
\begin{align}
\frac{d}{d\mbox{log}(\mu)}\alpha_i^{-1}(\mu)=
-\frac{b_i}{2\pi},
\label{Eq:RGE-gauge-coupling-alpha}
\end{align}
where $i$ stands for a gauge group $G$; e.g., $4C$ stands for the gauge
coupling constant of $SU(4)_C$, and the beta function coefficient  is
given by 
\begin{align}
b_i=
-\frac{11}{3}\sum_{\rm Vector}T(R_V)
+\frac{2}{3}\sum_{\rm Weyl}T(R_F)
+\frac{1}{6}\sum_{\rm Real}T(R_S),
\label{Eq:beta-function-coeff-general}
\end{align}
where Vector, Weyl, and Real stand for real vector, Weyl fermion, and
real scalar fields, respectively.
Since the vector bosons are gauge bosons, they belong to the
adjoint representation of the Lie group $G$: $T(R_V)=C_2(G)$.
$C_2(G)$ is the quadratic Casimir invariant of the adjoint
representation of $G$, and  
$T(R_i)$ is a Dynkin index of the irreducible representation $R_{i}$ of
$G$.
Note that when the Lie group $G$ is spontaneously broken into its
Lie subgroup $G'$, it is convenient to use the irreducible representations
of $G'$. (For the Dynkin index and the branching rules, see
e.g., Refs.~\cite{Yamatsu:2015gut,McKay:1981} or
calculated by using appropriate computer programs such as
Susyno \cite{Fonseca:2011sy}, LieART \cite{Feger:2012bs,Feger:2019tvk},
and GroupMath \cite{Fonseca:2020vke}.
For the RGEs at the two-loop level,  see, e.g.,
Refs.~\cite{Machacek:1983tz,Machacek:1983fi,Machacek:1984zw}.)

Let us consider the RGEs for gauge coupling constants in the pNGB DM model
shown in 
Tables~\ref{Tab:Matter_content-GUT-fermion},  
\ref{Tab:Matter_content-GUT-scalar}, and
\ref{Tab:Matter_content-GUT-gauge}.
For the energy scale between $M_{Z}<\mu<M_{I}$ and
$M_I<\mu<M_U$, we use the RGEs for the gauge coupling constants of
$G_{\rm SM}$ and $G_{\rm PS}$, respectively. In the following calculation,
we assume that there is only one intermediate scale $M_I$ and one
unification scale $M_U$, which should be recognized as effective
scales.

We can obtain the beta function coefficients of the gauge coupling
constants of $G_{\rm SM}$ and $G_{\rm PS}$ by using the generic RGE in 
Eq.~(\ref{Eq:beta-function-coeff-general}) and 
the matter content of the model given in
Tables~\ref{Tab:Matter_content-GUT-fermion},
\ref{Tab:Matter_content-GUT-scalar}, and 
\ref{Tab:Matter_content-GUT-gauge}.
The beta function coefficients of $G_{\rm SM}$
in $M_{Z}<\mu<M_{I}$ are given by
\begin{align}
\left(
\begin{array}{c}
b_{3C}\\
b_{2L}\\
b_{1Y}\\
\end{array}
 \right)
=\left(
\begin{array}{c}
-7\\
-19/6\\
+41/10\\
\end{array}
 \right),
\label{Eq:b_SM} 
\end{align}
where $i=3C,2L,1Y$ stand for $SU(3)_C$, $SU(2)_L$, $U(1)_Y$, respectively,
and we took the $SU(5)$ normalization for $U(1)_Y$.
(The values of $b_i$ are the same as the ordinary SM.)
The beta function coefficients of $G_{\rm PS}$ in $M_{I}<\mu<M_{U}$ are
given by
\begin{align}
\left(
\begin{array}{c}
b_{4C}\\
b_{2L}'\\
b_{2R}\\
\end{array}
 \right)
 =\left(
\begin{array}{c}
 -{22}/{3}\\
 -3\\
 +13/3\\
\end{array}
 \right),
\label{Eq:b_PS}
\end{align}
where $i=4C,2L,2R$ stand for $SU(4)_C$, $SU(2)_L$, $SU(2)_R$,
respectively.
To distinguish the beta function coefficient of the $SU(2)_L$ in
$G_{\rm SM}$ and that in $G_{\rm PS}$, we use unprimed and primed, and
the same notation is used below.

To solve the above RGEs, we need to set the initial conditions at
$\mu=M_Z$. The gauge coupling constants must satisfy 
the matching conditions between $G_{\rm SM}$ and $G_{\rm PS}$ at $\mu=M_I$
and also the matching condition between $G_{\rm PS}$ and $SO(10)$ at
$\mu=M_U$. They are listed below.
\begin{itemize}
\item  The input parameters for the three SM gauge coupling
       constants at $\mu=M_Z=91.1876\pm 0.0021$$\,$GeV
       are given in Ref.~\cite{Zyla:2020zbs}:
\begin{align}
\alpha_{3C}(M_Z)=0.1181 \pm 0.0011,\ \ 
\alpha_{2L}(M_Z)=\frac{\alpha_{\rm EM}(M_{Z})}{\sin^2\theta_W(M_Z)},\ \
\alpha_{1Y}(M_Z)=\frac{5\alpha_{\rm EM}(M_{Z})}{3\cos^2\theta_W(M_Z)},
\label{Eq:Alpha-exp-1}
\end{align}
       where the experimental values of the EM gauge coupling constant
       $\alpha_{\rm EM}$ and the Weinberg angle are given as
\begin{align}
\alpha_{\rm EM}^{-1}(M_{Z})=127.955\pm 0.010,\ \ 
\sin^2\theta_W(M_Z)=0.23122\pm 0.00003.
\label{Eq:Alpha-exp-2}
\end{align}

\item 
The matching conditions between $G_{\rm SM}$ and $G_{\rm PS}$ at
$\mu=M_I$ are given by
\begin{align}
\alpha_{3C}(M_{I})=\alpha_{4C}(M_{I}),\ \
\alpha_{2L}(M_{I})=\alpha^\prime_{2L}(M_{I}),\ \
\alpha^{-1}_{1Y}(M_{I})&=
 \frac{3}{5}\alpha^{-1}_{2R}(M_{I})
 +\frac{2}{5}\alpha^{-1}_{4C}(M_{I}),
\end{align}
where they are determined by the normalization conditions of the
      generators of $G_{\rm PS}$ and $G_{\rm SM}$.
(See e.g., Ref.~\cite{Mohapatra2002} at one-loop level;
Refs.~\cite{Deshpande:1992au,Deshpande:1992em} at two-loop level.)

\item 
The matching condition at the unification scale $M_{U}$ is
given by
\begin{align}
\alpha_{4C}(M_{U})=
\alpha_{2L}'(M_{U})=
\alpha_{2R}(M_{U}).
\end{align}
\end{itemize}

By using the RGEs of $G_{\rm SM}$ and $G_{\rm PS}$ and the matching
conditions at $\mu=M_I$ and $M_U$, we can obtain $M_I$ and $M_U$ as 
\begin{align}
 M_I&=M_Z\ \mbox{exp}\left[\frac{A_1B_3-A_3B_1}{A_2B_3-A_3B_2}\right]
 ,\nonumber\\
 M_U&
 =
 M_Z\ \mbox{exp}\left[\left(\frac{A_1B_3-A_3B_1}{A_2B_3-A_3B_2}\right)+
 \left(\frac{A_1B_2-A_2B_1}{A_3B_2-A_2B_3}\right)\right],
\label{Eq:MI-MU} 
\end{align}
where 
\begin{align}
 &A_1=\alpha_{3C}^{-1}(M_Z)-\alpha_{2L}^{-1}(M_Z),\
 A_2=\frac{b_{3C}-b_{2L}}{2\pi},\ 
 A_3=\frac{b_{4C}-b_{2L}'}{2\pi},\nonumber\\
 &B_1=\frac{5}{3}\left(
 \alpha_{3C}^{-1}(M_Z)-\alpha_{1Y}^{-1}(M_Z)\right),\ 
 B_2=\frac{5}{3}\frac{b_{3C}-b_{1Y}}{2\pi},\ 
 B_3=\frac{b_{4C}-b_{2R}}{2\pi}.
\end{align}
The gauge coupling constants such as 
$\alpha_{4C}(M_U)$ and $\alpha_{2L}'(M_U)$ are also expressed by
the $Z$ boson mass $M_Z$, the gauge coupling constants at $\mu=M_Z$ and
the beta function coefficients of $G_{\rm SM}$ and $G_{\rm PS}$ $b_i$s.
(The detail analysis is given in Appendix~\ref{Sec:RGE-gauge}.)

\begin{figure}[tbh]
\begin{center}
\includegraphics[bb=0 0 481 348,height=5.5cm]{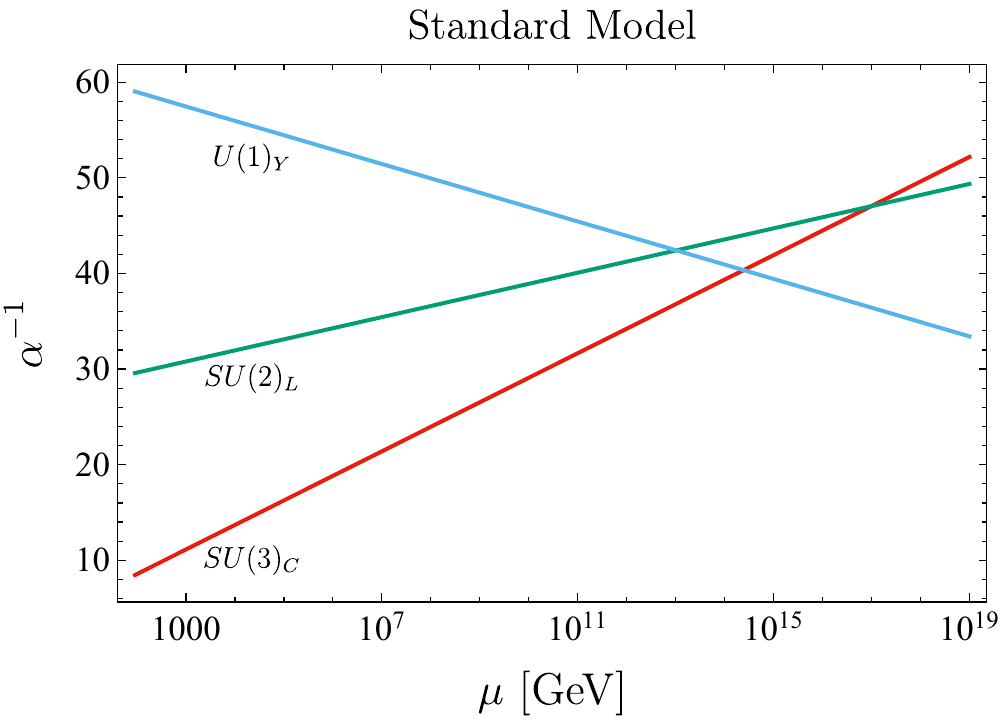}
\includegraphics[bb=0 0 400 290,height=5.5cm]{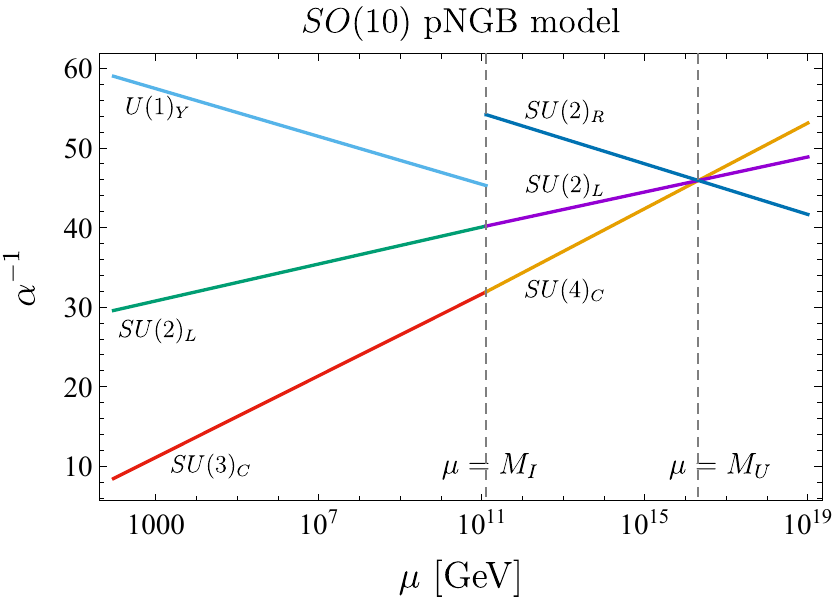}
\end{center}
\caption{\small The gauge coupling constants $\alpha_i$ vs the energy scale
 $\mu$ for the SM (the left figure) and the $SO(10)$ pNGB model (the
 right figure) are shown.
 The left figure shows the energy dependence of three gauge coupling
 constants of $SU(3)_C$,  $SU(2)_L$, and  $U(1)_Y$,
 $\alpha_{3C}$, $\alpha_{2L}$, and $\alpha_{1Y}$
 in all the energy ranges
 $\mu=[M_Z,M_H]$, where $M_H=10^{19}$$\,$GeV.
 The right figure shows $\alpha_{3C}$, $\alpha_{2L}$,
 and $\alpha_{1Y}$
 in the energy ranges  $\mu=[M_Z,M_I]$;
 $\alpha_{4C}$, $\alpha_{2L}$,  $\alpha_{2R}$ in the energy ranges
 $\mu=[M_I,M_H]$, where the value of 
 $\alpha_{3C}$ is fixed as the central value $\alpha_{3C}(M_Z)=0.1181$
 \cite{Zyla:2020zbs}.
}
\label{Figure:RGE-gauge-coupling}
\end{figure}

By substituting $b_i$ in Eqs.~(\ref{Eq:b_SM}) and (\ref{Eq:b_PS})
and the parameters at $\mu=M_Z$ in Eqs.~(\ref{Eq:Alpha-exp-1}) and
(\ref{Eq:Alpha-exp-2}) into 
the expressions of $M_I$ and $M_U$ in Eq.~(\ref{Eq:MI-MU}),
we find the values of the $M_I$ and $M_U$ as
\begin{align}
 M_I= (1.261\pm 0.242)\times 10^{11}\,\mbox{GeV},\ \ 
 M_U= (2.057\pm 0.688)\times 10^{16}\,\mbox{GeV}.
 \label{Eq:MI-MU-value}
\end{align}
Note that we ignore such as mass splitting at the intermediate and
unification scales, so the uncertainty must be larger.
The values of the model parameters at $\mu=M_I$ are given by
\begin{align}
&\alpha_{4C}^{-1}(M_I)=31.92\pm 0.23,\ \
\alpha_{2L}^{\prime-1}(M_I)=40.19\pm 0.10,\ \
\alpha_{2R}^{-1}(M_I)=54.20\pm 0.26.
\end{align}
We also find the gauge coupling constants of $U(1)_{B-L}$ and $U(1)_R$
at $\mu=M_I$
\begin{align}
 g_{B-L}(M_I)= 0.3843\pm 0.0009,\ \ 
 g_{R}(M_I)=0.4815\pm 0.0011,
\end{align}
by using $g_{B-L}(M_I)=\sqrt{\frac{3\pi}{2}\alpha_{4C}(M_I)}$ and
$g_{R}(M_I)=\sqrt{4\pi\alpha_{2R}(M_I)}$.
Since the standard normalization of $U(1)_{B-L}$ is not the same 
as that of ``$U(1)_{B-L}$''$(\subset SU(4)_{C}/SU(3)_C)$, the modified
normalization factor is used.
The unified gauge coupling constants at $\mu=M_U$ is given by
\begin{align}
 \alpha_U^{-1}=45.92\pm 0.50.
 \label{Eq:AlphaU-value}
\end{align}
The energy dependence of the gauge coupling constants $\alpha_i(\mu)$
in the $SO(10)$ pNGB model is plotted in
Fig.~\ref{Figure:RGE-gauge-coupling}.

As the same as the usual GUT models, nucleon can decay via the
so-called lepto-quark gauge bosons. The proton lifetime via
the gauge bosons is roughly
estimated as  $\tau\simeq{M_U^4}/{\alpha_U^2 m_p^5}$
\cite{Nath:2006ut,Mohapatra2002,Zyla:2020zbs},
where $m_p$ is the proton mass and the gauge boson masses are assumed to
be $M_U$.
From the values of $M_U$ and $\alpha_U$ given in
Eqs.~(\ref{Eq:MI-MU-value}) and 
(\ref{Eq:AlphaU-value}),
the proton lifetime
$\tau\simeq 1.1\times 10^{37}$ years is predicted.
It is far from the current constraint
$\tau(p\to e^+\pi^0)>2.4\times 10^{34}$ years at $90\%$ CL
\cite{Takenaka:2020vqy};
$M_U>(4.3-4.8)\times 10^{15}$\,GeV for
$40\lesssim\alpha_U^{-1}\lesssim50$. 
There is contribution for the proton decay modes via colored scalar 
fields shown in Table~\ref{Tab:Matter_content-GUT-scalar}.
The color triplet component of $\Phi_{\bf 10}$ has assumed to
have $\mathcal{O}(M_U)$, so the contribution for the proton decay 
via the Yukawa coupling constant
$\mathsf{y}_{\bf 10}^{(ab)}$ of the term
$\mathsf{\Phi}_{{\bf 10}}
\left(\mathsf{\Psi}_{\bf 16}^{(a)}\mathsf{\Psi}_{\bf 16}^{(b)}\right)_{\bf 10}$
in Eq.~(\ref{Eq:Lagrangian-SO10})
is small.
Color non-singlet components of $\mathsf{\Phi}_{\bf \overline{126}}$ have
assumed to $\mathcal{O}(M_I)$, so the contribution for the proton decay 
via the Yukawa coupling constant
$\mathsf{y}_{\bf \overline{126}}^{(ab)}$ of the term
$\mathsf{\Phi}_{\overline{\bf 126}}^*
\left(\mathsf{\Psi}_{\bf 16}^{(a)}\mathsf{\Psi}_{\bf 16}^{(b)}\right)_{\bf \overline{126}}$
in Eq.~(\ref{Eq:Lagrangian-SO10})
can be larger than the current experimental bounds.
This leads to an upper bound of the values of 
$\mathsf{y}_{\bf \overline{126}}^{(ab)}$ in the model.

We comment on proton decay via a colored Higgs scalar or lepto-quark 
scalar denoted as $S_1$ in Ref.~\cite{Dorsner:2016wpm}, which
belongs to $({\bf 3,1},1/3)$ under $G_{\rm SM}$.
In the following, we omit Clebsch-Gordan coefficients for simplicity.
When the lepto-quark scalar $S_1$ has di-quark and
quark-lepton couplings, there are proton decay modes such as
$p\to e^+\pi^0$, and the proton lifetime is roughly estimated as
$\tau\simeq m_{LQ}^4/(|y|^2|z|^2 m_p^5)$, where $m_{LQ}$ is a
lepto-quark mass, $y$ and $z$ represent generic values of relevant
Yukawa coupling constants of the lepto-quark with the quark-lepton and
quark-quark pairs, respectively.
For example, for the lepto-quark with the intermediate scale mass
$m_{LQ}=M_I$ and the universal Yukawa coupling constants $|y|=|z|$,
we obtain a constraint for the Yukawa coupling constants
$|y|=|z|\lesssim 4.2\times 10^{-6}$ from the current constraint
$\tau(p\to e^+\pi^0)>2.4\times 10^{34}$ years at $90\%$ CL.
To apply this for the current model,
for the scalar field $S_1$ in ${\bf 10}$ of $SO(10)$, which belongs to
${\bf (6,1,1)}$ under $G_{\rm PS}$,
the mass of the lepto-quark scalar is the unification scale mass
$m_{LQ}=M_U$ and the Yukawa coupling constants are roughly expected as 
$|y|=|z|\simeq|y_{\bf 10}^{(11)}|$.
The current constraint $\tau(p\to e^+\pi^0)>2.4\times 10^{34}$
years at $90\%$ CL leads to $|y_{\bf 10}^{(11)}|\lesssim 0.68$.
To realize the mass of up quark, $y_{\bf 10}^{(11)}$ is roughly 
$\mathcal{O}(10^{-5})$, so it is consistent with the current constraint,
where the actual values of the Yukawa coupling constants depend on how
to realized the observed quark and lepton masses.
Next, for the scalar fields $S_{1{\bf (\overline{10},1,3)}}$
and $S_{1{\bf (1,1,3)}}$ in ${\bf \overline{126}}$ of $SO(10)$,
which belongs to ${\bf (\overline{10},1,3)}$ and 
${\bf (6,1,1)}$ under $G_{\rm PS}$.
The lepto-quark scalar 
$S_{1{\bf (\overline{10},1,3)}}$ and $S_{1{\bf (6,1,1)}}$
have the intermediate scale mass $M_I$ and the unification scale mass
$M_U$, respectively.
For $S_{1{\bf (\overline{10},1,3)}}$, 
the Yukawa coupling couplings are given by
$|y|=0$ and $|z|\simeq|y_{\bf \overline{126}}|$,
so the proton decay mediated by $S_{1{\bf (\overline{10},1,3)}}$
does not occur. Therefore, this does not lead to any constraint for
$y_{\bf \overline{126}}^{(ab)}$.
For  $S_{1{\bf (6,1,1)}}$, the Yukawa coupling couplings are given by
$|y|=|z|\simeq|y_{\bf \overline{126}}|$.
the current constraint $\tau(p\to e^+\pi^0)>2.4\times 10^{34}$
years at $90\%$ CL leads to 
$|y_{\bf \overline{126}}^{(11)}|\lesssim 0.68$ as the same as
$S_1$ in ${\bf 10}$ of $SO(10)$.
In the above discussion, we assumed
$S_{1{\bf (\overline{10},1,3)}}$ does not mix with $S_{1{\bf (6,1,1)}}$,
but they have the same quantum numbers, so it depends on the structure
of the scalar potential, they can be mixed in general.
Even when the mixing parameter denoted as $\varepsilon$ between
$S_{1{\bf (\overline{10},1,3)}}$ and 
$S_{1{\bf (6,1,1)}}$ is about the ratio of the masses
$\varepsilon\simeq M_I/M_U\simeq 6.1\times 10^{-6}$,
the current constraint $\tau(p\to e^+\pi^0)>2.4\times 10^{34}$
years at $90\%$ CL leads to
the constraint for the first generation Yukawa coupling constant
$|y_{\bf \overline{126}}^{(11)}|\lesssim 1.7\times 10^{-3}$.
(For $\varepsilon=1$,
$|y_{\bf \overline{126}}^{(11)}|\lesssim 4.2\times 10^{-6}$.)

Further, we comment on the relation between neutrino masses and the
Yukawa coupling constants $y_{\bf \overline{126}}^{(ab)}$ of the cubic
term 
$\mathsf{\Phi}_{\bf 16}\mathsf{\Phi}_{\bf 16}\mathsf{\Phi}_{\bf \overline{126}}$.
Since the right-handed neutrino masses are given by
$M_{N}^{(ab)}=y_{\bf \overline{126}}^{(ab)}v_\phi$, 
we obtain $2.1\times 10^8\,\mbox{GeV}\lesssim
M_{N}^{(11)}=y_{\bf \overline{126}}^{(11)}v_\phi
\lesssim 1.4\times 10^{11}\,\mbox{GeV}$
for $1.7\times 10^{-3}\lesssim y_{\bf \overline{126}}^{(11)}\lesssim 0.68$
and $v_\phi=M_I$.
From the Type-I see-saw mechanism,
the light neutrino mass is roughly
$m_\nu^{(11)}\simeq |y_{\bf 10}^{(11)}v|^2/M_{N}^{(11)}$
when we ignore the off-diagonal part of $M_{N}^{(ab)}$.
Therefore,
$4.4\times 10^{-8}\,\mbox{eV}\lesssim m_\nu^{(11)}\lesssim
2.9\times 10^{-5}\,\mbox{eV}$
for $1.7\times 10^{-3}\lesssim y_{\bf \overline{126}}^{(11)}\lesssim 0.68$,
$|y_{\bf 10}^{(11)}|\simeq 10^{-5}$ and
$v\simeq 246\,\mbox{GeV}$.
The proton decay constraints only a part of
the Yukawa coupling constants $y_{\bf \overline{126}}^{(ab)}$, so
it is expected that the observed neutrino masses can be reproduced,
but to perform it properly, we need to investigate how to reproduce the
observed quark and charged lepton masses. We leave it for a future study.

Up to this point, we only consider the specific symmetry breaking
pattern, $SO(10)$ broken to $G_I=G_{\rm PS}$ at $\mu=M_U$ in
Eq.(\ref{Eq:Symmetry-breaking-pattern}).
We comment on other cases
$G_I=G_{\rm PS}\times D$, $G_{\rm LR}$, $G_{\rm LR}\times D$
discussed in e.g.,
Refs.~\cite{Deshpande:1992au,Deshpande:1992em,Babu:2015bna,Ferrari:2018rey}, 
where $D$ stands for a discrete $Z_2$ left-right exchange symmetry
\cite{Chang:1983fu,Chang:1984uy}.
(Note that the same analysis in $SO(10)$ GUT models whose matter content
is slightly different from the present model has been already
discussed in e.g., Refs.~\cite{Deshpande:1992au,Deshpande:1992em} by
using two-loop RGEs \cite{Jones:1981we} and the corresponding matching
condition \cite{Hall:1980kf,Chang:1984qr}.)
To realize the appropriate symmetry breaking patterns, we  need
different $SO(10)$ breaking Higgs fields; each  $G_I=G_{\rm PS}$,
$G_{\rm PS}\times D$, $G_{\rm LR}$, $G_{\rm LR}\times D$ is
realized by the VEV of a scalar field in  e.g., ${\bf 210}$, ${\bf 54}$,
${\bf 45}$, ${\bf 210}$ of $SO(10)$, respectively.

The values of $M_I$, $M_U$, and $\alpha_U^{-1}$ for several matter
contents and symmetry breaking patterns are summarized
in Table~\ref{Tab:Summary-RGEs-SO10}, which are estimated by using
each analytical solution shown in Appendix~\ref{Sec:RGE-gauge}.
Substituting the values of $M_U$ and $\alpha_U^{-1}$
for the $G_{\rm PS}\times D$ and $G_{\rm LR}\times D$ cases
into $\tau\simeq{M_U^4}/{\alpha_U^2 m_p^5}$,
rapid proton decay is expected.
For the $G_{\rm LR}$ case, the proton decay
via lept-quark gauge bosons is consistent with the current experimental
constraints, but the pNGB cannot be identified as DM because pNGB decays
too rapidly or the observed relic abundance cannot be reproduced.

\begin{table}[tbh]
\begin{center}
{\footnotesize
\begin{tabular}{|c|c|c|cc|c|}\hline
 \rowcolor[gray]{0.8}
 Group $G_I$&Scalars at $\mu=M_I$ & $b_j$&
 \multicolumn{2}{c|}{
 \begin{tabular}{cc}
  \multicolumn{2}{c}{$\mbox{log}_{10}(M/1\mbox{[GeV]})$}\\\hline
  $M_I$&$M_U$\\
 \end{tabular}
 }
 &$\alpha_U^{-1}$
 \\\hline
 $G_{\rm PS}$
 & \begin{tabular}{c}
 $({\bf 1,2,2})_{\bf 10}$\\
 {$({\bf \overline{4},1,2})_{\bf 16}$}\\
 $({\bf \overline{10},1,3})_{\bf \overline{126}}$\\
 \end{tabular}
 &
 $\left(
 \begin{array}{c}
 b_{4C}\\
 b_{2L}'\\
 b_{2R}\\
 \end{array}
 \right)=\left(
 \begin{array}{c}
 -\frac{22}{3}\\
 -3\\
 +\frac{13}{3}\\
 \end{array}
 \right)$
 &$11.10\pm 0.08$
 &$16.31\pm 0.15$
 &$45.92\pm 0.50$
 \\\hline
 $G_{\rm PS}\times D$
 & \begin{tabular}{c}
 $({\bf 1,2,2})_{\bf 10}$\\
 {$({\bf {4},2,1})_{\bf 16}$}\\
 {$({\bf \overline{4},1,2})_{\bf 16}$}\\
 $({\bf \overline{10},1,3})_{\bf \overline{126}}$\\
 $({\bf 10,3,1})_{\bf \overline{126}}$\\
  \end{tabular}
 &
 $\left(
 \begin{array}{c}
 b_{4C}\\
 b_{2L}'\\
 b_{2R}\\
 \end{array}
 \right)=\left(
 \begin{array}{c}
 -4\\
 +\frac{13}{3}\\
 +\frac{13}{3}\\
 \end{array}
 \right)$
 &$13.71\pm 0.03$
 &$15.22\pm 0.04$
 &$40.82\pm 0.13$
 \\\hline
 $G_{\rm LR}$
 & \begin{tabular}{c}
 $({\bf 1,2,2},0)_{\bf 10}$\\
 {$({\bf 1,1,2},1)_{\bf 16}$}\\
 $({\bf 1,1,3},2)_{\bf \overline{126}}$\\
   \end{tabular}
 &
 $\left(
 \begin{array}{c}
 b_{3C}'\\
 b_{2L}'\\
 b_{2R}\\
 b_{B-L}\\
 \end{array}
 \right)=\left(
 \begin{array}{c}
 -7\\
 -3\\
 -\frac{13}{6}\\
 +\frac{23}{4}\\
 \end{array}
 \right)$
 &$8.57\pm 0.06$
 &$16.64\pm 0.13$
 &$46.13\pm0.41$
 \\\hline
 $G_{\rm LR}\times D$
 & \begin{tabular}{c}
 $({\bf 1,2,2},0)_{\bf 10}$\\
 {$({\bf 1,1,2},1)_{\bf 16}$}\\
 {$({\bf 1,2,1},1)_{\bf 16}$}\\
 $({\bf 1,1,3},2)_{\bf \overline{126}}$\\
 $({\bf 1,3,1},-2)_{\bf \overline{126}}$\\
  \end{tabular}
 &
 $\left(
 \begin{array}{c}
 b_{3C}'\\
 b_{2L}'\\
 b_{2R}\\
 b_{B-L}\\
 \end{array}
 \right)=\left(
 \begin{array}{c}
 -7\\
 -\frac{13}{6}\\
 -\frac{13}{6}\\
 +\frac{15}{2}\\
 \end{array}
 \right)$
 &$10.11\pm 0.04$
 &$15.57\pm 0.09$
 &$43.38\pm 0.30$
 \\\hline
\end{tabular}
}
 \caption{\small The values of $M_I$, $M_U$, and $\alpha_U^{-1}$ for several
 matter contents and symmetry breaking patterns are summarized.
 The top of the table corresponds to the present $SO(10)$ pNGB model.
 The first, second, and third columns represent the intermediate scale
 group $G_I$, the matter content for scalar sector at $\mu=M_I$,
 the beta function coefficients $b_j$ of $G_I$, respectively.
 The fourth and fifth columns show the values of $M_I$, $M_U$, and
 $\alpha_U^{-1}$.
 The subscript in the second column stands for each $SO(10)$
 representation.
 }
\label{Tab:Summary-RGEs-SO10}
\end{center}  
\end{table}

\section{Long-lived pNGB as DM candidate}
\label{Sec:Dark-matter}

The DM lifetime should be longer than the age of the universe,
$10^{17}\,\mathrm{s}$ at least. 
The bound on DM lifetime becomes stronger depending on DM decay
channels due to the constraint of cosmic-ray observations.  
In particular, the bound from gamma-ray observations is strong as
roughly $\tau_{\chi}\gtrsim10^{27}\,\mathrm{s}$ for two body
decays \cite{Baring:2015sza}.  
Since the DM lifetime is proportional to the power of the VEV
$v_{\phi}$, it becomes longer for larger $v_\phi$.  
The evaluation of DM lifetime without GUT has been studied in
Refs.~\cite{Abe:2020iph,Okada:2020zxo}, 
and it has turned out that the VEV should roughly be
$v_{\phi}\gtrsim10^{13}\,\mathrm{GeV}$ in order to be consistent with  
the gamma-ray observations if three body decays
$\chi\to h_if\overline{f}$ and $Zf\overline{f}$ can occur.  
Since in the current GUT pNGB model the kinetic mixing $\sin\epsilon$
and the VEV $v_{\phi}$ are fixed to be $\sin\epsilon=-\sqrt{2/5}$ and  
$v_\phi\simeq 10^{11}\,\mathrm{GeV}$ by the requirement of the gauge
coupling unification, 
the three body decays should kinematically be forbidden. 
Therefore we consider the mass region
$m_{\chi}\lesssim\mathcal{O}(100)\,\mathrm{GeV}$ and estimate dominant
four body decay channels.

Before proceeding to four body decays, we comment on the two body decay
channel $\chi\to\nu\nu$, which is possible even in the case
$m_{\chi}\lesssim\mathcal{O}(100)\,\mathrm{GeV}$. 
Similarly to the $U(1)_{B-L}$ model in the previous  
paper~\cite{Abe:2020iph}, this process occurs via
the scalar mixing given by Eq.~(\ref{Eq:U-odd}) and 
the mixing between the  left-handed and right-handed neutrinos after the
electroweak symmetry breaking.
The decay width for this channel is calculated as
\begin{align}
 \Gamma_{\nu\nu}&=
 \frac{m_{\chi}}{64\pi}\frac{v_s^2}{v_{\phi}^4}\sum_{i}m_{\nu_i}^2
 \nonumber\\
&=5\times10^{-59}\,\mathrm{GeV}
\left(\frac{m_{\chi}}{100\,\mathrm{GeV}}\right)
\left(\frac{v_s}{1\,\mathrm{TeV}}\right)^2
\left(\frac{10^{11}\,\mathrm{GeV}}{v_{\phi}}\right)^4
\sum_{i}\left(\frac{m_{\nu_i}}{0.1\,\mathrm{eV}}\right)^2,
\label{eq:dmdecay}
\end{align}
where $m_{\nu_i}$ is the small neutrino mass eigenvalues. 
Eq.~(\ref{eq:dmdecay}) roughly corresponds to the lifetime
$\tau_{\nu\nu}=\mathcal{O}(10^{34})\,\mathrm{s}$, which is too small to
be observed in neutrino cosmic-rays
\cite{PalomaresRuiz:2007ry,Covi:2009xn} 
because of the suppression by the small neutrino mass squared
$m_{\nu_i}^2$.
Note that since the scale of the VEV in the GUT pNGB model is $v_{\phi}\simeq 10^{11}\,\mathrm{GeV}$ which is much smaller than the previous analysis \cite{Abe:2020iph}, the order of the lifetime for this channel is much shorter. However it is still too long to be detectable by experiments and observations.

\begin{figure}[tbh]
\begin{center}
\includegraphics[bb=0 0 402 302,height=4.5cm]{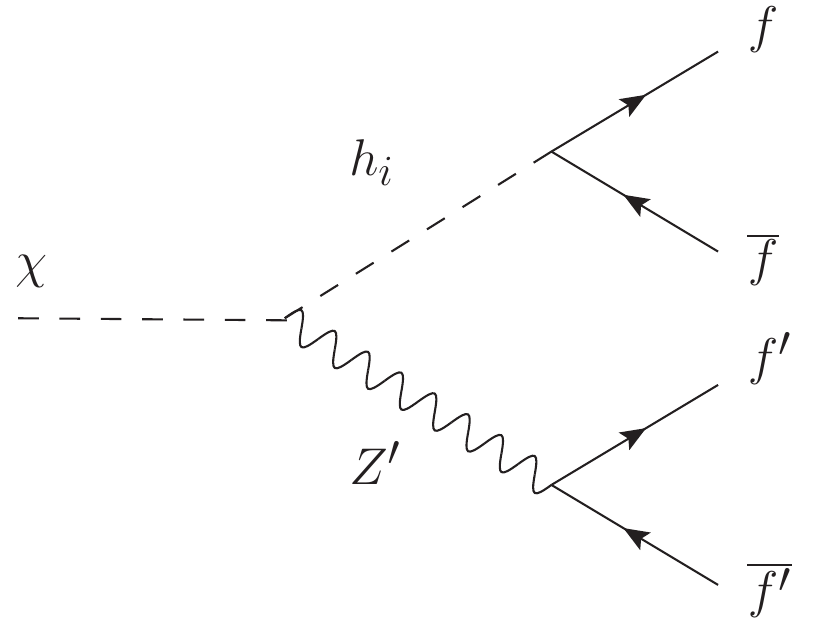}\ \
\includegraphics[bb=0 0 402 302,height=4.5cm]{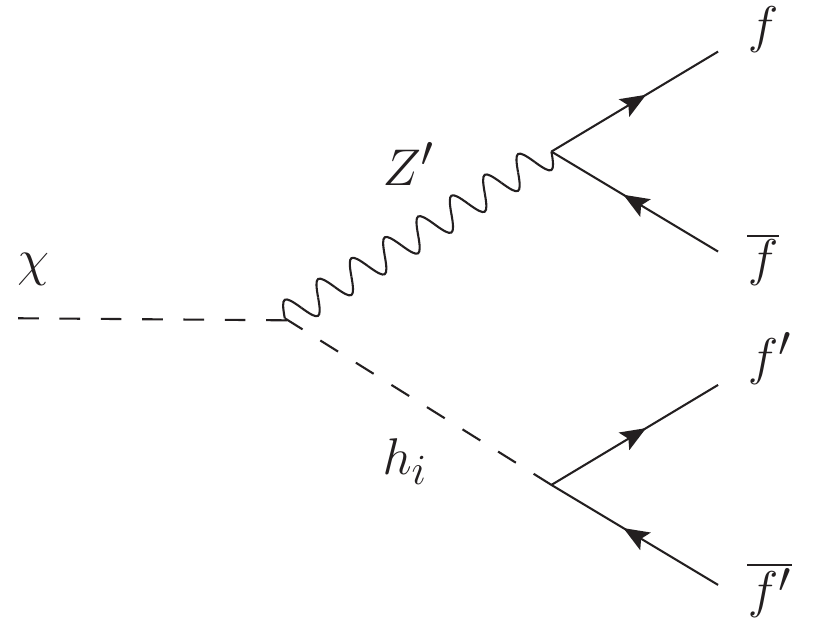}
\end{center}
\caption{\small The Feynman diagrams  for the four body
decays $\chi\to f\bar{f}f'\bar{f'}$ are shown.
}
\label{Figure:DM-4-body-decay}
\end{figure}

\begin{figure}[htb]
\begin{center}
\includegraphics[bb=0 0 286 207,scale=0.775]{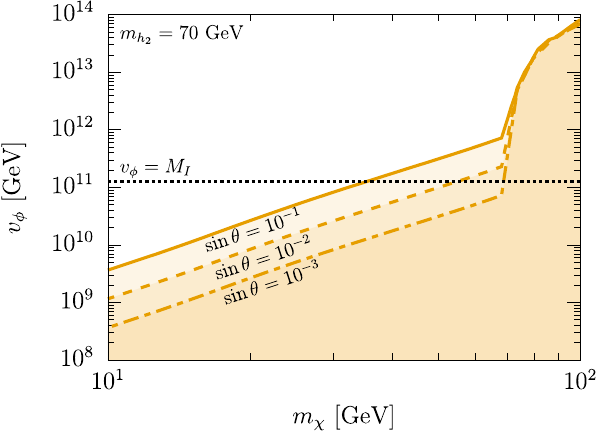}
\includegraphics[bb=0 0 286 207,scale=0.775]{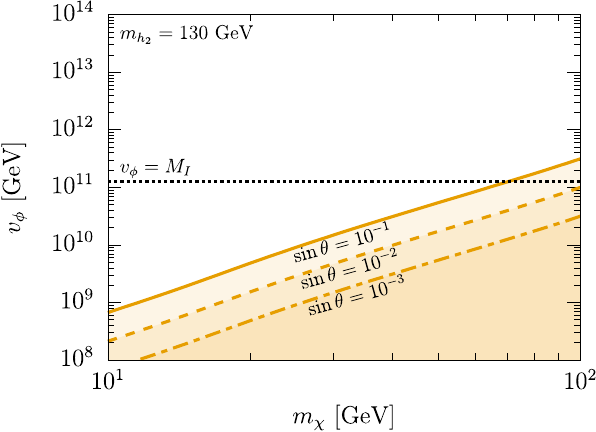}
\caption{\small Parameter space in the ($m_\chi$, $v_\phi$) plane where the
 second Higgs mass is fixed to be $m_{h_2}=70\,\mathrm{GeV}$ in the left
 and $130\,\mathrm{GeV}$ in the right.  
The orange region is excluded by the bound of the gamma-ray observations
 ($\tau_\chi=10^{27}\,\mathrm{s}$) for $\sin\theta=10^{-1},10^{-2}$ and
 $10^{-3}$.} 
\label{fig:decay}
\end{center}
\end{figure}

The four body decay processes
$\chi\to f\overline{f}f^\prime\overline{f^\prime}$ mediated by
$h_i,Z,Z^\prime$ can occur as shown in
Fig.~\ref{Figure:DM-4-body-decay}. 
Note that if $f$ and $f^\prime$ are identical particles, additional
diagrams exist due to interference. 
We numerically evaluated the decay width for all the four body decay
processes using CalcHEP~\cite{Belyaev:2012qa}, and furthermore we took
into account three body decay processes when these are kinematically
possible. The results are shown in Fig.~\ref{fig:decay} in ($m_\chi$,
$v_{\phi}$) plane where the second Higgs mass is fixed to be
$m_{h_2}=70\,\mathrm{GeV}$ (left) and $130\,\mathrm{GeV}$ (right). 
The orange region below the solid, dashed and dot-dashed lines are the
region where the DM lifetime is shorter than the conservative
bound $\tau_{\chi}=10^{27}\,\mathrm{s}$ 
for the Higgs mixing angle $\sin\theta=10^{-1},10^{-2},10^{-3}$,
respectively.\footnote{The actual bound on the DM lifetime for
four body decays is weaker than $\tau_{\chi}\gtrsim10^{27}\,\mathrm{s}$
since the energy of the emitted gamma rays is softer than two body
decays.}  
The horizontal black dotted line denotes
$v_{\phi}=M_I=10^{11.10}\,\mathrm{GeV}$.
The most part of the region in the plots is dominated by the four body
decays except for the region $m_{\chi}\gtrsim 60\,\mathrm{GeV}$ in the
left panel where the three body decay $\chi\to h_2f\overline{f}$ can
open up. One can read off the upper bound of the DM mass
$m_{\chi}$ for a given mixing angle $\sin\theta$.  

\begin{figure}[htb]
\begin{center}
\includegraphics[bb=0 0 286 207,scale=0.775]{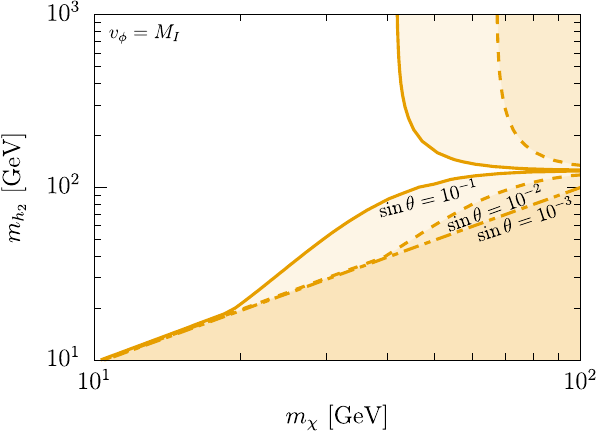}
 \caption{\small Parameter space in ($m_\chi$, $h_{h_2}$) plane, where the VEV
 is fixed to be $v_{\phi}=M_I$.  
 The orange region is excluded by the bound of the gamma-ray
 observations ($\tau_\chi=10^{27}\,\mathrm{s}$) for
 $\sin\theta=10^{-1},10^{-2}$ and $10^{-3}$.}
\label{fig:decay2}
\end{center}
\end{figure}

Fig.~\ref{fig:decay2} shows the parameter space in ($m_{\chi}$,
$m_{h_2}$) plane for the Higgs mixing angle $\sin\theta=10^{-1},10^{-2}$
and $10^{-3}$ where $v_\phi=M_I$. 
The region $m_{\chi}\gtrsim m_{h_2}$ is strongly constrained by three
body decay $\chi\to h_{2}f\overline{f}$ while the other region is
constrained by four body decays. 
In particular, if the second Higgs mass is degenerate with the SM-like
Higgs boson ($m_{h_1}\simeq m_{h_2}$), the four body decay width can be
small and the constraint is weaken. 
This is because the effective coupling $\chi$-$f$-$f^\prime$ mediated by
$h_1$ and $h_2$ becomes small when $m_{h_1}\simeq m_{h_2}$.

\begin{figure}[htb]
\begin{center}
\includegraphics[bb=0 0 286 207,scale=0.775]{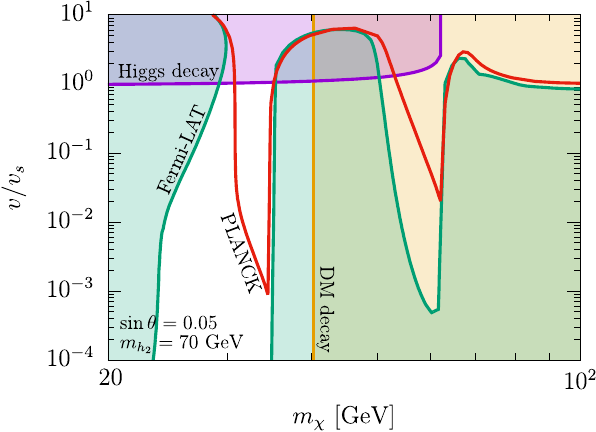}
\includegraphics[bb=0 0 286 207,scale=0.775]{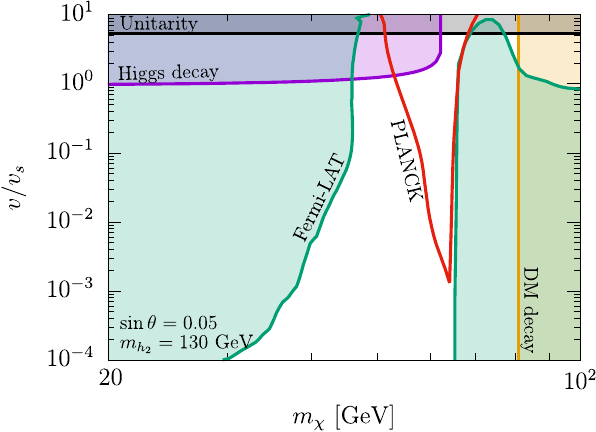}
\caption{\small Parameter space thermally reproducing the observed relic
 abundance consistent with some other observations.  
The red line represents the parameter space reproducing the correct
 thermal relic abundance $\Omega_{\chi} h^2\simeq 0.12$.  
The orange and green region are excluded by gamma-ray observations
 coming from the DM decay and annihilations, respectively.  
The purple region are excluded by the constraints of the Higgs invisible
 decay $h_{1}\to\chi\chi$ and the Higgs signal strength.  
The gray region is perturbative unitarity bound $\lambda_{S}>8\pi/3$.}
\label{fig:relic}
\end{center}
\end{figure}

Thermal relic abundance of DM is calculated using
micrOMEGAs~\cite{Belanger:2018ccd}.  
The results are shown in Fig.~\ref{fig:relic}, where the other
parameters are fixed to be $m_{h_2}=70\,\mathrm{GeV}$, $\sin\theta=0.05$
in the left panel and $m_{h_2}=130\,\mathrm{GeV}$ and $\sin\theta=0.05$
the right panel. The red line denotes the parameter space which can
reproduce the observed relic abundance of DM $\Omega_{\chi}
h^2\simeq 0.12$~\cite{Aghanim:2018eyx}. 
The purple region is excluded by the constraints of the Higgs invisible
decay and Higgs signal strength~\cite{Sirunyan:2018owy, Aaboud:2019rtt}, 
and the gray region is excluded by the perturbative unitarity bound
$\lambda_{S}<8\pi/3$ \cite{Chen:2014ask}. 
The green and orange region are ruled out by the constraints of the
gamma-ray observations for DM annihilations~\cite{Fermi-LAT:2016uux} and 
four body decays~\cite{Baring:2015sza}, respectively.
One can see that the thermal relic abundance can be consistent with all
the constraints when the DM mass is rather close to the
resonances $m_\chi\lesssim m_{h_i}/2$. This is the characteristic due to
the requirement from the gauge coupling unification in the current GUT
pNGB model. 

We comment on the allowed parameter space $m_{\chi}\lesssim m_{h_i}/2$. 
For the second Higgs mass rather heavier than the SM-like Higgs mass,  
the constraint of the gamma-ray observations can be avoided only if the  
DM mass is light enough $m_{\chi}\lesssim 35\,\mathrm{GeV}$ as can be  
seen from Fig.~\ref{fig:decay2}. On the other hand, this mass region  
cannot be consistent with the thermal relic abundance of DM since it is  
far from the Higgs resonances. Therefore the mass region $m_{h_2}\gtrsim  
m_{h_1}$ is completely excluded as long as thermal production mechanism  
of DM is assumed.
For more precise calculations in the region $m_{\chi}\lesssim
m_{h_i}/2$, the effect of the early kinetic decoupling from the SM
thermal bath should be taken into account
\cite{Binder:2017rgn,Abe:2020obo}. 
If this effect is included, one can expect that the red line in
Fig.~\ref{fig:decay} is shifted slightly upward.

\section{Summary}
\label{Sec:Summary}

In this paper, we proposed an $SO(10)$ pNGB DM model in the framework of 
GUTs. Each Weyl fermion in ${\bf 16}$ of $SO(10)$ contains one
generation of quark and leptons. The SM Higgs and two complex scalar
fields $H$, $S$ and $\Phi$ in the previous
gauged $U(1)_{B-L}$ pNGB DM model are
embedded into scalar fields in ${\bf 10}$, ${\bf 16}$, and
${\bf \overline{126}}$ of $SO(10)$. 
Assuming a symmetry breaking pattern of $SO(10)$ to $G_{\rm PS}$ at 
$\mu=M_U$, and further to $G_{\rm SM}$ at $\mu=M_I$, 
the intermediate and unified scales $M_I$ and $M_U$,
the gauge coupling constants of $U(1)_{B-L}$,
and the kinetic mixing parameter of between $U(1)_Y$ and $U(1)_{B-L}$
are determined by solving the RGEs with appropriate matching
conditions such as gauge coupling unification at $\mu=M_U$.

The DM lifetime without GUT has analyzed in
Refs.~\cite{Abe:2020iph,Okada:2020zxo}. It suggests that the VEV should
roughly be the VEV of $\Phi$ $v_{\phi}\gtrsim10^{13}\,\mathrm{GeV}$ in
order to be consistent with the gamma-ray observations if three body
decays $\chi\to h_if\overline{f}$ and $Zf\overline{f}$ are possible.
In the current GUT pNGB model, the kinetic mixing and the
VEV are fixed to be $\sin\epsilon=-\sqrt{2/5}$ and
$v_\phi\simeq10^{11}\,\mathrm{GeV}$, respectively.
To satisfy the constraint from the
gamma-ray observations, the pNGB DM mass must be 
$m_{\chi}\lesssim\mathcal{O}(100)\,\mathrm{GeV}$
to forbid the three body decays kinematically.
In the mass region, the dominant contribution for DM decay channels
comes from four body decay channels $\chi\to f\bar{f}f'\bar{f'}$.
We find that the thermal relic abundance can be consistent with all the
constraints when the DM mass is rather close to the resonances
$m_\chi\lesssim m_{h_i}/2$.

\section*{Acknowledgments}
\noindent

This work was supported in part
by the MEXT Grant-in-Aid for Scientific Research on Innovation Areas
Grant No. JP18H05543 (K.T. and N.Y.)
and 
JSPS Grant-in-Aid for Scientific Research KAKENHI Grant Nos. 
JP20J11901 (Y.A.), JP20K22349 (T.T.), and JP19K23440 (N.Y.).
Numerical computation in this work was carried out at the Yukawa
Institute Computer Facility. 

\appendix

\section{Kinetic mixing as mass mixing}
\label{Sec:Kinetic-mixing}

As discussed in the main part of this paper,
the gauge kinetic mixing in Refs.~\cite{Abe:2020iph,Okada:2020zxo} is regarded as the
mixing angle. In this appendix, we will show this explicitly.
The scalar fields in Refs.~\cite{Abe:2020iph,Okada:2020zxo} are embedded into the
scalars of $SO(10)$ shown in Table~\ref{Tab:Matter_content-GUT-scalar}
as 
\begin{align}
 & \sPhi_{\bm{10}} \supset \phi_{(\bm{1}, \bm{2}, \bm{2})} \supset \phi_{(\bm{1}, \bm{2}, 1/2)} = H,
 \label{eq:HinSO(10)}
 \\
 & \sPhi_{\bm{16}} \supset \phi_{(\ol{\bm{4}}, \bm{1}, \bm{2})} \supset \phi_{(\bm{1}(+3), \bm{1}, -1/2)} = S,
 \label{eq:SinSO(10)}
 \\
 & \sPhi_{\ol{\bm{126}}} \supset \phi_{(\ol{\bm{10}}, \bm{1}, \bm{3})} \supset \phi_{(\bm{1}(+6), \bm{1}, -1)} = \Phi.
 \label{eq:PhiinSO(10)}
\end{align}
Here we will consider the following two symmetry breaking pattern:
\begin{align}
 G_{\mathrm{PS}} \to G_{\mathrm{SM}},
 \quad
 G_{\mathrm{PS}} \to G_{\mathrm{LR}} \to G_{\mathrm{SM}}.
\end{align}

\subsection{$G_{\mr{PS}} \to G_{\mr{SM}}$}

First, let us consider the following symmetry breaking pattern
\begin{align}
 SU(4)_C \times SU(2)_R \utxtarrow{\braket{\phi_{(\ol{\bm{10}}, \bm{1}, \bm{3})}} \neq 0,~
 \braket{\phi_{(\ol{\bm{4}}, \bm{1}, \bm{2})}} \neq 0 }
 SU(3)_C \times U(1)_Y,
\end{align}
using minimal scalar fields
Eqs.~\eqref{eq:HinSO(10)}--\eqref{eq:PhiinSO(10)}. 
This breaking pattern is suitable for the pNGB dark matter model
embedding into an $SO(10)$ GUT model because the intermediate scale can
be large enough to make the dark matter candidate long-lived. 

The covariant derivative of $G_{\mathrm{PS}}$ gauge group acts on $S$
and $\Phi$ as 
\begin{align}
 D_\mu S &= \del_\mu S 
 + i g_{4} G'^{\ol{\bm{3}},a}_\mu I^{SU(4)_C}_{\ol{\bm{3}}(-4),a} S
 + i g_{B-L} E_\mu Q^S_{B-L} S
 + i \frac{g_R}{\sqrt{2}} W'^+_\mu I^{SU(2)_R}_+ S
 + i g_R W'^3_\mu I^{SU(2)_R}_3 S
 \nonumber\\
 &= \del_\mu S 
 + i g_{4} G'^{\ol{\bm{3}},a}_\mu I^{SU(4)_C}_{\ol{\bm{3}}(-4),a} S
 + i \frac{g_R}{\sqrt{2}} W'^+_\mu I^{SU(2)_R}_+ S
 + i g_{B-L} E_\mu S 
 - \frac{i g_R}{2} W'^3_\mu S,
 \\
 D_\mu \Phi &= \del_\mu \Phi
 + i g_{4} G'^{\bm{3},a}_\mu I^{SU(4)_C}_{\bm{3}(4),a} \Phi
 + i g_{B-L} E_\mu Q^\Phi_{B-L} \Phi
 + i \frac{g_R}{\sqrt{2}} W'^+_\mu I^{SU(2)_R}_+ \Phi
 + i g_{R} W'^3_\mu I^{SU(2)_R}_{3} \Phi
 \nonumber\\
 &=\del_\mu \Phi 
 + i g_{4} G'^{\bm{3},a}_\mu I^{SU(4)_C}_{\bm{3}(4),a} \Phi
 + i \frac{g_R}{\sqrt{2}} W'^+_\mu I^{SU(2)_R}_+ \Phi
 + 2 i g_{B-L} E_\mu \Phi
 - i g_{R} W'^3_\mu \Phi
 ,
\end{align}
where $E_\mu$ is the gauge field associated with
$U(1)_{B-L} \subset SU(4)_C$ and $g_{B-L}$ is the gauge coupling
constant given by $ g_{B-L} = \sqrt{ \frac{3}{8}} g_4$. 
The $B-L$ charge comes from the diagonal component of $SU(4)$
denoted by
\begin{align}
 Q_{B-L} = \sqrt{ \frac{8}{3}} I^{SU(4)_C}_{15},
 \quad
 I^{SU(4)_C}_{15} = \sqrt{ \frac{3}{8}} \diag( 1/3, 1/3, 1/3, -1).
 \label{eq:defofQB-L}
\end{align}
$G'^{\bm{3},a}_\mu$ and $G'^{\ol{\bm{3}},a}_\mu$ are color charged
vector boson with the representation $\bm{3}(4)$ and $\ol{\bm{3}}(-4)$
of $SU(3)_C \times U(1)_{B-L}$ belonging to $\bm{15}$ of $SU(4)_C$
respectively. 
(For the details of the branching rules and the tensor products, see Ref.~\cite{Yamatsu:2015gut}.)
These scalars are assumed to develop the following VEVs,
\begin{align}
 \braket{S} = \frac{v_s}{\sqrt{2}},
 \quad
 \braket{\Phi} = \frac{v_\phi}{\sqrt{2}},
 \label{eq:VEVSPhi}
\end{align}
and these gives the mass terms of the gauge fields
\begin{align}
 \mc{L}_{SU(4)_C \times SU(2)_R, \text{mass}} =&
 G'^{\bm{3},a\dagger}_\mu M^2_{\bm{3}, ab} G'^{\bm{3},b \mu}
 + G'^{\ol{\bm{3}},a \dagger}_\mu M^2_{\ol{\bm{3}},ab} G'^{\ol{\bm{3}},b \mu} 
 + \frac{g_R^2}{4} (v_s^2 + 2 v_\phi^2) W'^-_\mu W'^{+\mu}
 \nonumber \\
 &~+\frac{1}{2} \biggl( \frac{v_s^2}{4} + v_\phi^2 \biggr)
 \bigl(2 g_{B-L} E_\mu  - g_R W'^3_\mu \bigr)^2,
 \label{eq:gaugemassofPS}
\end{align}
where the mass matrices for the color charged vector bosons 
$G'^{\bm{3},a}_\mu$ and $G'^{\ol{\bm{3}},a}_\mu$ are defined by
\begin{align}
 M^2_{\bm{3},ab} = \frac{g_4^2 v_\phi^2}{2} \tr \Bigl[ 
 \bigl( I^{SU(4)_C}_{\bm{3}(4),a} \bigr)^\dagger
 I^{SU(4)_C}_{\bm{3}(4), b} \Bigr],
 \quad
 M^2_{\ol{\bm{3}},ab} = \frac{g_4^2 v_s^2}{2} \tr \Bigl[
 \bigl( I^{SU(4)_C}_{\ol{\bm{3}}(-4),a} \bigr)^\dagger
 I^{SU(4)_C}_{\ol{\bm{3}}(-4),b} \Bigr].
\end{align}
The last term of Eq.~\eqref{eq:gaugemassofPS} leads the mass mixing
between $U(1)_{B-L} \subset SU(4)_C$ and $U(1)_{R} \subset SU(2)_R$, 
and the massless direction becomes $U(1)_Y$ in the SM gauge group.
From this term,
the massive vector boson $C'_\mu$ and the orthogonal massless gauge
boson $B'_\mu$ are introduced by 
\begin{align}
 \pmat{B'_\mu \\ C'_\mu}
 = \pmat{ \cos \epsilon & \sin \epsilon\\
 - \sin \epsilon & \cos \epsilon}
 \pmat{W'^3_\mu \\ E_\mu},
 \label{eq:mixingBXWpC}
\end{align}
where the mixing angle is defined by
\begin{align}
 \sin \epsilon = \frac{g_R}{\sqrt{ g_R^2 + 4 g_{B-L}^2}},
 \quad
 \cos \epsilon = \frac{2 g_{B-L}}{\sqrt{ g_R^2 + 4 g_{B-L}^2}},
 \label{eq:mixingangleepsilon}
\end{align}
and the mass of $C'_\mu$ becomes
$M_{C'}^2 = (g_R^2 + 4 g_{B-L}^2) (v_s^2 /4 + v_\phi^2)$.
In this basis, the Lagrangian is
\begin{align}
 \mc{L} \supset
 - \frac{1}{4} W^{a}_{\mu\nu}W^{a\mu\nu}
 - \frac{1}{4} B_{\mu\nu}'B^{\prime\mu\nu}
 - \frac{1}{4} C_{\mu\nu}'C^{\prime\mu\nu}
 + \frac{1}{2} M_{C'}^2 C_\mu'C^{\prime\mu}
 \label{eq:LagWBX}
\end{align}
If the color charged vector bosons are dropped,
the covariant derivative is rewritten by using these bosons as
\begin{align}
 D_\mu \supset i g_1 B'_\mu + i g_{C'} C'_\mu \biggl( \frac{Q_{B-L}}{2}
 - \sin^2 \epsilon\, Q_Y \biggr), 
\end{align}
where the hypercharge is defined by
\begin{align}
 Q_Y = I^{SU(2)_R}_3 + \frac{Q_{B-L}}{2},
 \label{eq:defofU(1)Y}
\end{align}
and the couplings are given by
\begin{align}
 g_1 = \frac{2 g_R g_{B-L} }{\sqrt{ g_R^2 + 4 g_{B-L}^2}},
 \quad
 g_{C'} = \sqrt{ g_R^2 + 4 g_{B-L}^2}.
 \label{eq:defofcoupling}
\end{align}

\subsubsection*{Correspondence between the pNGB model \cite{Abe:2020iph,Okada:2020zxo} and the $SO(10)$ pNGB model}

We will discuss the kinetic mixing in the GUT model.
First, from Eq.~\eqref{eq:mixingBXWpC}, $B'_\mu$ is written by using $(W'^3_\mu, E_\mu)$ as
$B'_\mu = W'^3_\mu / \cos \epsilon + \sin \epsilon E_\mu / \cos \epsilon$,
and
the field redefinition by $\cos \epsilon$ leads the canonically normalized gauge kinetic terms.
The massive direction of broken $U(1)$ symmetry does not change in this rewriting.
Then Let us introduce new fields after the rescaling
by
\begin{align}
 \pmat{B_\mu' \\
 C_\mu'}
 = \pmat{1 & \sin \epsilon \\
 0 & \cos \epsilon}
 \pmat{B_\mu \\
 C_\mu},
 \quad
 \pmat{B_\mu \\
 C_\mu} = \pmat{1 & -\tan \epsilon \\
 0 & 1/\cos \epsilon}
 \pmat{B_\mu' \\
 C'_\mu},
 \label{eq:mixing2}
\end{align}
so that the massive direction does not change but the massless component
is replaced. 
The relation between $(W'^3_\mu, E_\mu)$ and $(B_\mu, C_\mu)$ is given
by 
\begin{align}
 \pmat{ W'^3_\mu \\
 E_\mu} = \pmat{\cos \epsilon & 0 \\
 \sin \epsilon & 1}
 \pmat{B_\mu \\
 C_\mu}.
\end{align}
The $U(1)_{B-L} \times U(1)_{R}$ gauge sector in the Lagrangian
\eqref{eq:LagWBX} is rewritten by using these fields as 
\begin{align}
 \mc{L} \supset - \frac{1}{4}W^a_{\mu\nu}W^{a\mu\nu}
 - \frac{1}{4} B_{\mu\nu}B^{\mu\nu}
 - \frac{1}{4} C_{\mu\nu}C^{\mu\nu}
 - \frac{\sin \epsilon}{2} B_{\mu\nu} C^{\mu\nu}
 + \frac{1}{2} M_C^2 C_\mu C^\mu,
 \label{eq:LagofATTinPS}
\end{align}
with $M_{C}^2 = g_{B-L}^2 ( v_s^2 + 4 v_\phi^2)$,
and the covariant derivative is given by
\begin{align}
 D_\mu \supset&~ i g_{B-L} E_\mu Q_{B-L} + i g_R W'^3_{\mu} I^{SU(2)_R}_3
  = i g_{B-L} C_\mu Q_{B-L} + i g_1 B_\mu Q_Y,
 \label{eq:DmuofATTinPS}
\end{align}
where
Eqs.~\eqref{eq:defofU(1)Y} and \eqref{eq:defofcoupling} are used.
Eqs.~\eqref{eq:LagofATTinPS} and \eqref{eq:DmuofATTinPS} are parts of the Lagrangian of the gauged $U(1)_{B-L}$ pNGB model,
and the gauge kinetic mixing is naturally regarded as the mixing angle coming from
the GUT inspired symmetry breaking.
The correspondence is summarized in Table~\ref{tab:correspondence}.

\begin{table}[t]
\centering
\begin{tabular}{c|c} \hline
Gauged $U(1)_{B-L}$ model \cite{Abe:2020iph} & pNGB in $SO(10)$ GUT \\
$G_{\mathrm{SM}} \times U(1)_{B-L}$ & $G_{\mathrm{PS}}$ \\
\hhline{=|=}
$Q_{Y}$ & $Q_Y = I_{3}^{SU(2)_R} + \frac{Q_{B-L}}{2}$ \\
$Q_{B-L}$ & $Q_{B-L} = \sqrt{\frac{8}{3}} I^{SU(4)_C}_{15}$ \\
\hline
$B_\mu$ & $B_\mu$ in Eq.~\eqref{eq:mixing2}\\
$\hat{B}_\mu$ & $B_\mu'$ in Eq.~\eqref{eq:mixingBXWpC} \\
$X_\mu$ & $C_\mu$ in Eq.~\eqref{eq:mixing2}\\
$\hat{X}_\mu$ & $C'_\mu$ in Eq.~\eqref{eq:mixingBXWpC}\\ 
\hline
$g_{1}$ & $g_{1} = 2 g_R g_{B-L} / \sqrt{ g_R^2 + 4 g_{B-L}^2}$ \\
$g_{B-L}$ & $g_{B-L} = \sqrt{ \frac{3}{8}} g_4$ \\
$g_{2}$ & $g_L$ \\
\hline
$D_\mu = \del_\mu + ig_s G^a_\mu I^{SU(3)_C}_a + ig_2 W^a_\mu I^{SU(2)_L}_a  $ & $D_\mu = \del_\mu + ig_s G^a_\mu I^{SU(3)_C}_a + ig_L W^a_\mu I^{SU(2)_L}_a $ \\
$+ ig_1 Q_Y B_\mu + i g_{B-L} Q_{B-L} X_\mu$& $+ ig_{1} Q_Y B_\mu + i g_{B-L} Q_{B-L} C_\mu$ \\
\hline
\multicolumn{2}{c}{kinetic mixing}\\
\hline
 gauge kinetic mixing of $B_\mu$ and $X_\mu$: $\epsilon$ & gauge kinetic mixing of $B_\mu$ and $C_\mu$: $\epsilon$ \\
 $=$ free parameter & $=$ mixing angle $\epsilon$ of $(W'^3_\mu, E_\mu) \mapsto (B'_\mu, C'_\mu)$
 \\
 &in Eq.~\eqref{eq:LagWBX}\\
\hline
\end{tabular}
\caption{\small
The correspondence table of the kinetic mixing and the gauge fields 
between the gauged $U(1)_{B-L}$ model \cite{Abe:2020iph,Okada:2020zxo} 
and $SO(10)$ GUT model. 
}
\label{tab:correspondence}
\end{table}

\subsection{$G_{\mr{PS}} \to G_{\mr{LR}} \to G_{\mr{SM}}$}

If the adjoint Higgs bosons $\phi_{(\bm{15},\bm{1},\bm{1})}$ and
$\phi_{(\bm{1}, \bm{1}, \bm{3})}$ are introduced 
in addition to the scalars
Eqs.~\eqref{eq:HinSO(10)}--\eqref{eq:PhiinSO(10)}, these VEVs break the
Pati-Salam gauge symmetry as 
\begin{align}
 SU(4)_C \utxtarrow{\braket{\phi_{(\bm{15}, \bm{1}, \bm{1})}} \neq 0}
 SU(3)_C \times U(1)_{B-L}, 
 \quad
 SU(2)_R \utxtarrow{\braket{\phi_{(\bm{1}, \bm{1}, \bm{3})}} \neq 0}
 U(1)_{R}. 
\end{align}
By this breaking pattern, the covariant derivative of $G_{\mathrm{PS}}$
reduces to that of 
$SU(3)_C \times SU(2)_L \times U(1)_{R3} \times U(1)_{B-L}$ as
\begin{align}
 D_\mu = \del_\mu 
 + ig_s G^a_\mu I^{SU(3)_C}_a 
 + ig_{B-L} E_\mu Q_{B-L} + ig_L W^a_\mu I^{SU(2)_L}_a + i g_R W'^3_\mu
 I^{SU(2)_R}_3,  
 \label{eq:DGSMxU1BL}
\end{align}
where the $B-L$ charge is defined by
Eq.~\eqref{eq:defofQB-L} and the gauge couplings are introduced by
$g_s = g_4$, $g_C = \sqrt{\frac{3}{8}} g_4$. 
The VEVs of $S$ and $\phi$ \eqref{eq:VEVSPhi} break the residual gauge
symmetry as 
\begin{align}
 SU(3)_C \times SU(2)_L \times U(1)_{R} \times U(1)_{B-L} \to
 G_{\mathrm{SM}}, 
\end{align}
and lead the mass term for the gauge bosons
\begin{align}
 \mc{L}_{U(1)_{R3} \times U(1)_{B-L},\text{mass}} = \frac{1}{2}
 \biggl( \frac{v_s^2}{4} + v_\phi^2 \biggr)
 \bigl( 2 g_{B-L} E_\mu - g_R W'^3_\mu \bigr)^2,
\end{align}
which is same to the last term of Eq.~\eqref{eq:gaugemassofPS}.
In this breaking pattern, the charged gauge bosons become massive via
the VEV of the adjoint Higgs fields. 

The mixing angle $\epsilon$ and correspondence between the mixing angle
and kinetic mixing are same in the previous discussions.

\section{RGEs for gauge coupling constants}
\label{Sec:RGE-gauge}

Here we analyze the RGEs for gauge coupling constants of $G_{\rm SM}$ and
$G_I=G_{\rm PS},G_{\rm LR}$, and $SO(10)$ in the pNGB DM
model. (For the RGE analysis, see e.g., Ref.~\cite{Mohapatra2002}.)

The RGE for the gauge coupling constants given in
Eq.~(\ref{Eq:RGE-gauge-coupling-alpha}) can be solve as 
\begin{align}
\alpha_i^{-1}(\mu_1)=
\alpha_i^{-1}(\mu_0)-\frac{b_i}{2\pi}\log\left(\frac{\mu_1}{\mu_0}\right).
\label{Eq:RGE-solution}
\end{align}
when the beta function coefficients $b_i$ are constant in the energy
range $\mu_0<\mu<\mu_1$.
In the following, we apply the solution for $G_{\rm SM}$, and
$G_I=G_{\rm PS},G_{\rm LR}$ cases.

In the following, we find the intermediate scale $M_I$ and $M_U$ can be
described by using the gauge coupling constants of $G_{\rm SM}$ at
$\mu=M_Z$ and the beta function coefficients of $G_{\rm SM}$ and
$G_I(=G_{\rm PS},G_{\rm LR})$.
Therefore, all the gauge coupling constants such as the unified gauge
coupling constant $\alpha_U$ can be analytically solved if they exist.

\subsection{$G_I=G_{\rm PS}$ case}
\label{Sec:GPS}

We list up the RGEs of $G_{\rm SM}$ and $G_{\rm PS}$ in
$M_Z<\mu<M_I$ and $M_I<\mu<M_U$, respectively, and the matching 
conditions at $\mu=M_I,M_U$.

\subsubsection{$M_Z<\mu< M_I$}

For $M_Z<\mu< M_I$, the RGEs of the gauge coupling constants of
$G_{\rm SM}=SU(3)_C\times SU(2)_L\times U(1)_Y$ are given by 
\begin{align}
\alpha_{3C}^{-1}(\mu)&=
\alpha_{3C}^{-1}(M_Z)-\frac{b_{3C}}{2\pi}\log\left(\frac{\mu}{M_Z}\right),\allowdisplaybreaks[1]\nonumber\\
\alpha_{2L}^{-1}(\mu)&=
\alpha_{2L}^{-1}(M_Z)-\frac{b_{2L}}{2\pi}\log\left(\frac{\mu}{M_Z}\right),\allowdisplaybreaks[1]\nonumber\\
\alpha_{1Y}^{-1}(\mu)&=
\alpha_{1Y}^{-1}(M_Z)-\frac{b_{1Y}}{2\pi}\log\left(\frac{\mu}{M_Z}\right).
\end{align}

\subsubsection{$\mu=M_I$}

The matching conditions between $G_{\rm SM}$ and
$G_{\rm PS}=SU(4)_C\times SU(2)_L\times SU(2)_R$ at $\mu=M_I$
are given as
\begin{align}
\alpha_{4C}^{-1}(M_I)&=
\alpha_{3C}^{-1}(M_I),\allowdisplaybreaks[1]\nonumber\\ 
\alpha_{2L}^{\prime-1}(M_I)&=
\alpha_{2L}^{-1}(M_I),\allowdisplaybreaks[1]\nonumber\\ 
\alpha_{2R}^{-1}(M_I)&=
\frac{5}{3}\alpha_{1Y}^{-1}(M_I)
-\frac{2}{3}\alpha_{3C}^{-1}(M_I).
\end{align}

\subsubsection{$M_I<\mu< M_U$}

For $M_I<\mu< M_U$, the RGEs of the gauge coupling constants of
$G_{\rm PS}$  are given by 
\begin{align}
\alpha_{4C}^{-1}(\mu)&=
\alpha_{4C}^{-1}(M_I)-\frac{b_{4C}}{2\pi}\log\left(\frac{\mu}{M_I}\right)
 =\alpha_{3C}^{-1}(M_Z)-\frac{b_{3C}}{2\pi}\log\left(\frac{M_I}{M_Z}\right)-\frac{b_{4C}}{2\pi}\log\left(\frac{\mu}{M_I}\right)
 ,\allowdisplaybreaks[1]\nonumber\\
\alpha_{2L}^{\prime-1}(\mu)&=
\alpha_{2L}^{\prime-1}(M_I)-\frac{b_{2L}'}{2\pi}\log\left(\frac{\mu}{M_I}\right)
 =\alpha_{2L}^{-1}(M_Z)-\frac{b_{2L}}{2\pi}\log\left(\frac{M_I}{M_Z}\right)
-\frac{b_{2L}'}{2\pi}\log\left(\frac{\mu}{M_I}\right),\allowdisplaybreaks[1]\nonumber\\
\alpha_{2R}^{-1}(\mu)&=
\alpha_{2R}^{-1}(M_I)-\frac{b_{2R}}{2\pi}\log\left(\frac{\mu}{M_I}\right)\nonumber\\
&=
 \frac{5}{3}\alpha_{1Y}^{-1}(M_Z)
-\frac{2}{3}\alpha_{3C}^{-1}(M_Z)
-\left(\frac{5}{3}\frac{b_{1Y}}{2\pi}-\frac{2}{3}\frac{b_{3C}}{2\pi}\right)
 \log\left(\frac{M_I}{M_Z}\right) 
-\frac{b_{2R}}{2\pi}\log\left(\frac{\mu}{M_I}\right).
\end{align}

\subsubsection{$\mu= M_U$}

For $\mu=M_U$, the matching condition
between $G_{\rm PS}$ and $SO(10)$ at $\mu=M_U$
is given by
\begin{align}
\alpha_{4C}^{-1}(M_U)=
\alpha_{2L}^{\prime-1}(M_U)=
\alpha_{2R}^{-1}(M_U),
\label{Eq:Matching-condition-MU} 
\end{align}
where 
\begin{align}
\alpha_{4C}^{-1}(M_U)&=
\alpha_{3C}^{-1}(M_Z)-\frac{b_{3C}}{2\pi}\log\left(\frac{M_I}{M_Z}\right)-\frac{b_{4C}}{2\pi}\log\left(\frac{M_U}{M_I}\right)
 ,\allowdisplaybreaks[1]\nonumber\\
\alpha_{2L}^{\prime-1}(M_U)&=
\alpha_{2L}^{-1}(M_Z)-\frac{b_{2L}}{2\pi}\log\left(\frac{M_I}{M_Z}\right)
-\frac{b_{2L}'}{2\pi}\log\left(\frac{M_U}{M_I}\right),\allowdisplaybreaks[1]\nonumber\\
\alpha_{2R}^{-1}(M_U)&=
 \frac{5}{3}\alpha_{1Y}^{-1}(M_Z)
-\frac{2}{3}\alpha_{3C}^{-1}(M_Z)
-\left(\frac{5}{3}\frac{b_{1Y}}{2\pi}-\frac{2}{3}\frac{b_{3C}}{2\pi}\right)
 \log\left(\frac{M_I}{M_Z}\right) 
-\frac{b_{2R}}{2\pi}\log\left(\frac{M_U}{M_I}\right).
\end{align}

\subsubsection{$M_I$ and $M_U$}

From the matching condition in Eq.~(\ref{Eq:Matching-condition-MU}),
we can analytically solve the intermediate scale $M_I$
and unification scale $M_U$ as 
\begin{align}
 M_I&=M_Z\ \mbox{exp}\left[\frac{A_1B_3-A_3B_1}{A_2B_3-A_3B_2}\right]
 ,\nonumber\\
 M_U&
 =
 M_Z\ \mbox{exp}\left[\left(\frac{A_1B_3-A_3B_1}{A_2B_3-A_3B_2}\right)+
 \left(\frac{A_1B_2-A_2B_1}{A_3B_2-A_2B_3}\right)\right],
\end{align}
where 
\begin{align}
 &A_1=\alpha_{3C}^{-1}(M_Z)-\alpha_{2L}^{-1}(M_Z),\
 A_2=\frac{b_{3C}-b_{2L}}{2\pi},\ 
 A_3=\frac{b_{4C}-b_{2L}'}{2\pi},\nonumber\\
 &B_1=\frac{5}{3}\left(
 \alpha_{3C}^{-1}(M_Z)-\alpha_{1Y}^{-1}(M_Z)\right),\ 
 B_2=\frac{5}{3}\frac{b_{3C}-b_{1Y}}{2\pi},\ 
 B_3=\frac{b_{4C}-b_{2R}}{2\pi}.
\end{align}

\subsection{$G_I=G_{\rm LR}$ case}
\label{Sec:GLR}

We list up the RGEs of $G_{\rm SM}$ and $G_I=G_{\rm LR}$ in
$M_Z<\mu<M_I$ and $M_I<\mu<M_U$, respectively, and the
matching conditions at $\mu=M_I,M_U$.

\subsubsection{$M_Z<\mu< M_I$}

For $M_Z<\mu< M_I$, the RGEs of the gauge coupling constants of
$G_{\rm SM}=SU(3)_C\times SU(2)_L\times U(1)_Y$ are given by 
\begin{align}
\alpha_{3C}^{-1}(\mu)&=
\alpha_{3C}^{-1}(M_Z)-\frac{b_{3C}}{2\pi}\log\left(\frac{\mu}{M_Z}\right),\allowdisplaybreaks[1]\nonumber\\
\alpha_{2L}^{-1}(\mu)&=
\alpha_{2L}^{-1}(M_Z)-\frac{b_{2L}}{2\pi}\log\left(\frac{\mu}{M_Z}\right),\allowdisplaybreaks[1]\nonumber\\
\alpha_{1Y}^{-1}(\mu)&=
\alpha_{1Y}^{-1}(M_Z)-\frac{b_{1Y}}{2\pi}\log\left(\frac{\mu}{M_Z}\right).
\end{align}

\subsubsection{$\mu=M_I$}

The matching conditions between $G_{\rm SM}$ and
$G_{\rm LR}=SU(3)_C\times SU(2)_L\times SU(2)_R\times U(1)_{B-L}$ at
$\mu=M_I$ are given as
\begin{align}
\alpha_{3C}^{\prime -1}(M_I)&=
\alpha_{3C}^{-1}(M_I),\allowdisplaybreaks[1]\nonumber\\ 
\alpha_{2L}^{\prime-1}(M_I)&=
\alpha_{2L}^{-1}(M_I),\allowdisplaybreaks[1]\nonumber\\ 
\alpha_{2R}^{-1}(M_I)&=
\frac{5}{3}\alpha_{1Y}^{-1}(M_I)
-\frac{2}{3}\alpha_{B-L}^{-1}(M_I).
\end{align}
Note that unlike the above $G_I=G_{\rm PS}$ case,
the gauge coupling constants of $G_{\rm LR}$ at $\mu=M_I$ 
cannot be determined only by using those of $G_{\rm SM}$ at $\mu=M_I$.
To fix them, we need to use the matching conditions of the gauge
coupling constants at $\mu=M_U$.

\subsubsection{$M_I<\mu< M_U$}

For $M_I<\mu< M_U$, the RGEs of the gauge coupling constants of
$G_{\rm LR}$  are given by 
\begin{align}
\alpha_{3C}^{\prime-1}(\mu)&=
\alpha_{3C}^{\prime-1}(M_I)-\frac{b_{3C}'}{2\pi}\log\left(\frac{\mu}{M_I}\right)
 =\alpha_{3C}^{-1}(M_Z)-\frac{b_{3C}}{2\pi}\log\left(\frac{M_I}{M_Z}\right)-\frac{b_{3C}'}{2\pi}\log\left(\frac{\mu}{M_I}\right)
 ,\allowdisplaybreaks[1]\nonumber\\
\alpha_{2L}^{\prime-1}(\mu)&=
 \alpha_{2L}^{\prime-1}(M_I)-\frac{b_{2L}'}{2\pi}\log\left(\frac{\mu}{M_I}\right)
 =\alpha_{2L}^{-1}(M_Z)-\frac{b_{2L}}{2\pi}\log\left(\frac{M_I}{M_Z}\right)
-\frac{b_{2L}'}{2\pi}\log\left(\frac{\mu}{M_I}\right),\allowdisplaybreaks[1]\nonumber\\
\alpha_{2R}^{-1}(\mu)&=
\alpha_{2R}^{-1}(M_I)-\frac{b_{2R}}{2\pi}\log\left(\frac{\mu}{M_I}\right),\allowdisplaybreaks[1]\nonumber\\
\alpha_{B-L}^{-1}(\mu)&=
\alpha_{B-L}^{-1}(M_I)-\frac{b_{B-L}}{2\pi}\log\left(\frac{\mu}{M_I}\right).
\end{align}

\subsubsection{$\mu= M_U$}

For $\mu=M_U$, the matching condition
between $G_{\rm LR}$ and $SO(10)$ at $\mu=M_U$
is given by
\begin{align}
\alpha_{3C}^{\prime-1}(M_U)=
\alpha_{2L}^{\prime-1}(M_U)=
\alpha_{2R}^{-1}(M_U)=
\alpha_{B-L}^{-1}(M_U),
\label{Eq:Matching-condition-MU-GLR}
\end{align}
where 
\begin{align}
\alpha_{3C}^{\prime-1}(M_U)&=
\alpha_{3C}^{-1}(M_Z)-\frac{b_{3C}}{2\pi}\log\left(\frac{M_I}{M_Z}\right)-\frac{b_{3C}'}{2\pi}\log\left(\frac{M_U}{M_I}\right)
 ,\allowdisplaybreaks[1]\nonumber\\
\alpha_{2L}^{\prime-1}(M_U)&=
\alpha_{2L}^{-1}(M_Z)-\frac{b_{2L}}{2\pi}\log\left(\frac{M_I}{M_Z}\right)
-\frac{b_{2L}'}{2\pi}\log\left(\frac{M_U}{M_I}\right),\allowdisplaybreaks[1]\nonumber\\
\alpha_{2R}^{-1}(M_U)&=
 \frac{5}{3}\alpha_{1Y}^{-1}(M_Z)
-\frac{2}{3}\alpha_{B-L}^{-1}(M_I)
-\frac{5}{3}\frac{b_{1Y}}{2\pi}
 \log\left(\frac{M_I}{M_Z}\right) 
-\frac{b_{2R}}{2\pi}\log\left(\frac{M_U}{M_I}\right),\allowdisplaybreaks[1]\nonumber\\
\alpha_{B-L}^{-1}(M_U)&=
 \alpha_{B-L}^{-1}(M_I)
-\frac{b_{B-L}}{2\pi}\log\left(\frac{M_U}{M_I}\right).
\end{align}

\subsubsection{$M_I$ and $M_U$}

From the matching condition in Eq.~(\ref{Eq:Matching-condition-MU-GLR}),
we can analytically solve the intermediate scale $M_I$
and unification scale $M_U$ as 
\begin{align}
 M_I&=M_Z\ \mbox{exp}\left[\frac{C_1D_3-C_3D_1}{C_2D_3-C_3D_2}\right]
 ,\allowdisplaybreaks[1]\nonumber\\
 M_U&
 =
 M_Z\ \mbox{exp}\left[\left(\frac{C_1D_3-C_3D_1}{C_2D_3-C_3D_2}\right)+
 \left(\frac{C_1D_2-C_2D_1}{C_3D_2-C_2D_3}\right)\right],
\end{align}
where 
\begin{align}
 &C_1=\alpha_{3C}^{-1}(M_Z)-\alpha_{2L}^{-1}(M_Z),\
 C_2=\frac{b_{3C}-b_{2L}}{2\pi},\ 
 C_3=\frac{b_{3C}'-b_{2L}'}{2\pi},\allowdisplaybreaks[1]\nonumber\\
 &D_1=
 \alpha_{2L}^{-1}(M_Z)-\alpha_{1Y}^{-1}(M_Z),\ 
 D_2=\frac{b_{2L}-b_{1Y}}{2\pi},\ 
 D_3=\frac{b_{2L}'-\frac{3b_{2R}+2b_{B-L}}{5}}{2\pi}.
\end{align}

\bibliographystyle{utphys} 
\bibliography{../../../arxiv/reference}

\providecommand{\href}[2]{#2}\begingroup\raggedright\begin{thebibliography}{10}

\bibitem{Corbelli:1999af}
E.~Corbelli and P.~Salucci, ``{The Extended Rotation Curve and the Dark Matter
  Halo of M33},'' \href{http://dx.doi.org/10.1046/j.1365-8711.2000.03075.x}{{
  Mon. Not. Roy. Astron. Soc.} {\bfseries 311} (2000) 441--447},
  \href{http://arxiv.org/abs/astro-ph/9909252}{{\ttfamily
  arXiv:astro-ph/9909252}}.

\bibitem{Sofue:2000jx}
Y.~Sofue and V.~Rubin, ``{Rotation Curves of Spiral Galaxies},''
  \href{http://dx.doi.org/10.1146/annurev.astro.39.1.137}{{ Ann. Rev. Astron.
  Astrophys.} {\bfseries 39} (2001) 137--174},
  \href{http://arxiv.org/abs/astro-ph/0010594}{{\ttfamily
  arXiv:astro-ph/0010594}}.

\bibitem{Massey:2010hh}
R.~Massey, T.~Kitching, and J.~Richard, ``{The Dark Matter of Gravitational
  Lensing},'' \href{http://dx.doi.org/10.1088/0034-4885/73/8/086901}{{ Rept.
  Prog. Phys.} {\bfseries 73} (2010) 086901},
  \href{http://arxiv.org/abs/1001.1739}{{\ttfamily arXiv:1001.1739
  [astro-ph.CO]}}.

\bibitem{Aghanim:2018eyx}
{\bfseries Planck} Collaboration, N.~Aghanim {et al.}, ``{Planck 2018 Results.
  VI. Cosmological Parameters},''
  \href{http://dx.doi.org/10.1051/0004-6361/201833910}{{ Astron. Astrophys.}
  {\bfseries 641} (2020) A6}, \href{http://arxiv.org/abs/1807.06209}{{\ttfamily
  arXiv:1807.06209 [astro-ph.CO]}}.

\bibitem{Randall:2007ph}
S.~W. Randall, M.~Markevitch, D.~Clowe, A.~H. Gonzalez, and M.~Bradac,
  ``{Constraints on the Self-Interaction Cross-Section of Dark Matter from
  Numerical Simulations of the Merging Galaxy Cluster 1E 0657-56},''
  \href{http://dx.doi.org/10.1086/587859}{{ Astrophys. J.} {\bfseries 679}
  (2008) 1173--1180}, \href{http://arxiv.org/abs/0704.0261}{{\ttfamily
  arXiv:0704.0261 [astro-ph]}}.

\bibitem{Freytsis:2010ne}
M.~Freytsis and Z.~Ligeti, ``{On Dark Matter Models with Uniquely
  Spin-Dependent Detection Possibilities},''
  \href{http://dx.doi.org/10.1103/PhysRevD.83.115009}{{ Phys. Rev. D}
  {\bfseries 83} (2011) 115009},
  \href{http://arxiv.org/abs/1012.5317}{{\ttfamily arXiv:1012.5317 [hep-ph]}}.

\bibitem{Ipek:2014gua}
S.~Ipek, D.~McKeen, and A.~E. Nelson, ``{A Renormalizable Model for the
  Galactic Center Gamma Ray Excess from Dark Matter Annihilation},''
  \href{http://dx.doi.org/10.1103/PhysRevD.90.055021}{{ Phys. Rev. D}
  {\bfseries 90} no.~5, (2014) 055021},
  \href{http://arxiv.org/abs/1404.3716}{{\ttfamily arXiv:1404.3716 [hep-ph]}}.

\bibitem{Arcadi:2017wqi}
G.~Arcadi, M.~Lindner, F.~S. Queiroz, W.~Rodejohann, and S.~Vogl,
  ``{Pseudoscalar Mediators: A WIMP Model at the Neutrino Floor},''
  \href{http://dx.doi.org/10.1088/1475-7516/2018/03/042}{{ JCAP} {\bfseries 03}
  (2018) 042}, \href{http://arxiv.org/abs/1711.02110}{{\ttfamily
  arXiv:1711.02110 [hep-ph]}}.

\bibitem{Sanderson:2018lmj}
N.~F. Bell, G.~Busoni, and I.~W. Sanderson, ``{Loop Effects in Direct
  Detection},'' \href{http://dx.doi.org/10.1088/1475-7516/2018/08/017}{{ JCAP}
  {\bfseries 08} (2018) 017}, \href{http://arxiv.org/abs/1803.01574}{{\ttfamily
  arXiv:1803.01574 [hep-ph]}}. [Erratum: JCAP 01, E01 (2019)].

\bibitem{Abe:2018emu}
T.~Abe, M.~Fujiwara, and J.~Hisano, ``{Loop Corrections to Dark Matter Direct
  Detection in a Pseudoscalar Mediator Dark Matter Model},''
  \href{http://dx.doi.org/10.1007/JHEP02(2019)028}{{ JHEP} {\bfseries 02}
  (2019) 028}, \href{http://arxiv.org/abs/1810.01039}{{\ttfamily
  arXiv:1810.01039 [hep-ph]}}.

\bibitem{Abe:2019wjw}
T.~Abe, M.~Fujiwara, J.~Hisano, and Y.~Shoji, ``{Maximum Value of the
  Spin-Independent Cross Section in the 2HDM+a},''
  \href{http://dx.doi.org/10.1007/JHEP01(2020)114}{{ JHEP} {\bfseries 01}
  (2020) 114}, \href{http://arxiv.org/abs/1910.09771}{{\ttfamily
  arXiv:1910.09771 [hep-ph]}}.

\bibitem{Barger:2010yn}
V.~Barger, M.~McCaskey, and G.~Shaughnessy, ``{Complex Scalar Dark Matter
  vis-a-vis CoGeNT, DAMA/LIBRA and XENON100},''
  \href{http://dx.doi.org/10.1103/PhysRevD.82.035019}{{ Phys. Rev. D}
  {\bfseries 82} (2010) 035019},
  \href{http://arxiv.org/abs/1005.3328}{{\ttfamily arXiv:1005.3328 [hep-ph]}}.

\bibitem{Gross:2017dan}
C.~Gross, O.~Lebedev, and T.~Toma, ``{Cancellation Mechanism for
  Dark-Matter--Nucleon Interaction},''
  \href{http://dx.doi.org/10.1103/PhysRevLett.119.191801}{{ Phys. Rev. Lett.}
  {\bfseries 119} no.~19, (2017) 191801},
  \href{http://arxiv.org/abs/1708.02253}{{\ttfamily arXiv:1708.02253
  [hep-ph]}}.

\bibitem{Ishiwata:2018sdi}
K.~Ishiwata and T.~Toma, ``{Probing Pseudo Nambu-Goldstone Boson Dark Matter at
  Loop Level},'' \href{http://dx.doi.org/10.1007/JHEP12(2018)089}{{ JHEP}
  {\bfseries 12} (2018) 089}, \href{http://arxiv.org/abs/1810.08139}{{\ttfamily
  arXiv:1810.08139 [hep-ph]}}.

\bibitem{Huitu:2018gbc}
K.~Huitu, N.~Koivunen, O.~Lebedev, S.~Mondal, and T.~Toma, ``{Probing
  Pseudo-Goldstone Dark Matter at the LHC},''
  \href{http://dx.doi.org/10.1103/PhysRevD.100.015009}{{ Phys. Rev. D}
  {\bfseries 100} no.~1, (2019) 015009},
  \href{http://arxiv.org/abs/1812.05952}{{\ttfamily arXiv:1812.05952
  [hep-ph]}}.

\bibitem{Cline:2019okt}
J.~M. Cline and T.~Toma, ``{Pseudo-Goldstone Dark Matter Confronts Cosmic Ray
  and Collider Anomalies},''
  \href{http://dx.doi.org/10.1103/PhysRevD.100.035023}{{ Phys. Rev. D}
  {\bfseries 100} no.~3, (2019) 035023},
  \href{http://arxiv.org/abs/1906.02175}{{\ttfamily arXiv:1906.02175
  [hep-ph]}}.

\bibitem{Jiang:2019soj}
X.-M. Jiang, C.~Cai, Z.-H. Yu, Y.-P. Zeng, and H.-H. Zhang,
  ``{Pseudo-Nambu-Goldstone Dark Matter and Two-Higgs-Doublet Models},''
  \href{http://dx.doi.org/10.1103/PhysRevD.100.075011}{{ Phys. Rev. D}
  {\bfseries 100} no.~7, (2019) 075011},
  \href{http://arxiv.org/abs/1907.09684}{{\ttfamily arXiv:1907.09684
  [hep-ph]}}.

\bibitem{Arina:2019tib}
C.~Arina, A.~Beniwal, C.~Degrande, J.~Heisig, and A.~Scaffidi, ``{Global Fit of
  Pseudo-Nambu-Goldstone Dark Matter},''
  \href{http://dx.doi.org/10.1007/JHEP04(2020)015}{{ JHEP} {\bfseries 04}
  (2020) 015}, \href{http://arxiv.org/abs/1912.04008}{{\ttfamily
  arXiv:1912.04008 [hep-ph]}}.

\bibitem{Abe:2020iph}
Y.~Abe, T.~Toma, and K.~Tsumura, ``{Pseudo-Nambu-Goldstone Dark Matter from
  Gauged $U(1)_{B-L}$ Symmetry},''
  \href{http://dx.doi.org/10.1007/JHEP05(2020)057}{{ JHEP} {\bfseries 05}
  (2020) 057}, \href{http://arxiv.org/abs/2001.03954}{{\ttfamily
  arXiv:2001.03954 [hep-ph]}}.

\bibitem{Okada:2020zxo}
N.~Okada, D.~Raut, and Q.~Shafi, ``{Pseudo-Goldstone Dark Matter in a Gauged
  $B-L$ Extended Standard Model},''
  \href{http://dx.doi.org/10.1103/PhysRevD.103.055024}{{ Phys. Rev. D}
  {\bfseries 103} no.~5, (2021) 055024},
  \href{http://arxiv.org/abs/2001.05910}{{\ttfamily arXiv:2001.05910
  [hep-ph]}}.

\bibitem{Zhang:2021alu}
Z.~Zhang, C.~Cai, X.-M. Jiang, Y.-L. Tang, Z.-H. Yu, and H.-H. Zhang, ``{Phase
  Transition Gravitational Waves from Pseudo-Nambu-Goldstone Dark Matter and
  Two Higgs Doublets},'' \href{http://dx.doi.org/10.1007/JHEP05(2021)160}{{
  JHEP} {\bfseries 05} (2021) 160},
  \href{http://arxiv.org/abs/2102.01588}{{\ttfamily arXiv:2102.01588
  [hep-ph]}}.

\bibitem{Georgi:1974sy}
H.~Georgi and S.~L. Glashow, ``{Unity of All Elementary Particle Forces},''
\href{http://dx.doi.org/10.1103/PhysRevLett.32.438}{{ Phys. Rev. Lett.}
  {\bfseries 32} (1974) 438--441}.

\bibitem{Heeck:2019kgr}
J.~Heeck and V.~Takhistov, ``{Inclusive Nucleon Decay Searches as a Frontier of
  Baryon Number Violation},''
  \href{http://dx.doi.org/10.1103/PhysRevD.101.015005}{{ Phys. Rev. D}
  {\bfseries 101} no.~1, (2020) 015005},
  \href{http://arxiv.org/abs/1910.07647}{{\ttfamily arXiv:1910.07647
  [hep-ph]}}.

\bibitem{Minkowski:1977sc}
P.~Minkowski, ``{$\mu \to e\gamma$ at a Rate of One Out of $10^{9}$ Muon
  Decays?},'' \href{http://dx.doi.org/10.1016/0370-2693(77)90435-X}{{ Phys.
  Lett. B} {\bfseries 67} (1977) 421--428}.

\bibitem{Fritzsch:1974nn}
H.~Fritzsch and P.~Minkowski, ``{Unified Interactions of Leptons and
  Hadrons},''
\href{http://dx.doi.org/10.1016/0003-4916(75)90211-0}{{ Ann. Phys.} {\bfseries
  93} (1975) 193--266}.

\bibitem{Pati:1974yy}
J.~C. Pati and A.~Salam, ``{Lepton Number as the Fourth Color},''
\href{http://dx.doi.org/10.1103/PhysRevD.10.275}{{ Phys. Rev.} {\bfseries D10}
  (1974) 275--289}.

\bibitem{Pati:1975ca}
J.~Pati, A.~Salam, and J.~Strathdee, ``{On Fermion number and its
  conservation},'' \href{http://dx.doi.org/10.1007/BF02849600}{{ Nuovo Cim. A}
  {\bfseries 26} (1975) 72--83}.

\bibitem{Mohapatra:1978fy}
R.~N. Mohapatra and G.~Senjanovic, ``{Natural Suppression of Strong P and T
  Noninvariance},''
\href{http://dx.doi.org/10.1016/0370-2693(78)90243-5}{{ Phys. Lett.} {\bfseries
  B79} (1978) 283--286}.

\bibitem{Slansky:1981yr}
R.~Slansky, ``{Group Theory for Unified Model Building},''
\href{http://dx.doi.org/10.1016/0370-1573(81)90092-2}{{ Phys. Rept.} {\bfseries
  79} (1981) 1--128}.

\bibitem{Yamatsu:2015gut}
N.~Yamatsu, ``{Finite-Dimensional Lie Algebras and Their Representations for
  Unified Model Building},''
\href{http://arxiv.org/abs/1511.08771}{{\ttfamily arXiv:1511.08771 [hep-ph]}}.

\bibitem{Bajc:2005zf}
B.~Bajc, A.~Melfo, G.~Senjanovic, and F.~Vissani, ``{Yukawa Sector in
  Non-Supersymmetric Renormalizable $SO(10)$},''
  \href{http://dx.doi.org/10.1103/PhysRevD.73.055001}{{ Phys. Rev. D}
  {\bfseries 73} (2006) 055001},
  \href{http://arxiv.org/abs/hep-ph/0510139}{{\ttfamily arXiv:hep-ph/0510139}}.

\bibitem{Aulakh:1982sw}
C.~Aulakh and R.~N. Mohapatra, ``{Implications of Supersymmetric $SO(10)$ Grand
  Unification},'' \href{http://dx.doi.org/10.1103/PhysRevD.28.217}{{ Phys. Rev.
  D} {\bfseries 28} (1983) 217}.

\bibitem{Babu:1992ia}
K.~Babu and R.~Mohapatra, ``{Predictive Neutrino Spectrum in Minimal $SO(10)$
  Grand Unification},'' \href{http://dx.doi.org/10.1103/PhysRevLett.70.2845}{{
  Phys. Rev. Lett.} {\bfseries 70} (1993) 2845--2848},
  \href{http://arxiv.org/abs/hep-ph/9209215}{{\ttfamily arXiv:hep-ph/9209215}}.

\bibitem{Aulakh:2003kg}
C.~S. Aulakh, B.~Bajc, A.~Melfo, G.~Senjanovic, and F.~Vissani, ``{The Minimal
  Supersymmetric Grand Unified theory},''
  \href{http://dx.doi.org/10.1016/j.physletb.2004.03.031}{{ Phys. Lett. B}
  {\bfseries 588} (2004) 196--202},
  \href{http://arxiv.org/abs/hep-ph/0306242}{{\ttfamily arXiv:hep-ph/0306242}}.

\bibitem{Fukuyama2005}
T.~Fukuyama, A.~Ilakovac, T.~Kikuchi, S.~Meljanac, and N.~Okada, ``{$SO(10)$
  Group Theory for the Unified Model Building},''
  \href{http://dx.doi.org/10.1063/1.1847709}{{ J. Math. Phys.} {\bfseries 46}
  (2005) 033505},
\href{http://arxiv.org/abs/hep-ph/0405300}{{\ttfamily arXiv:hep-ph/0405300}}.

\bibitem{Bertolini:2009qj}
S.~Bertolini, L.~Di~Luzio, and M.~Malinsky, ``{Intermediate Mass Scales in the
  Non-Supersymmetric $SO(10)$ Grand Unification: A Reappraisal},''
  \href{http://dx.doi.org/10.1103/PhysRevD.80.015013}{{ Phys. Rev.} {\bfseries
  D80} (2009) 015013},
\href{http://arxiv.org/abs/0903.4049}{{\ttfamily arXiv:0903.4049 [hep-ph]}}.

\bibitem{Altarelli:2013aqa}
G.~Altarelli and D.~Meloni, ``{A Non Supersymmetric SO(10) Grand Unified Model
  for All the Physics Below $M_{GUT}$},''
  \href{http://dx.doi.org/10.1007/JHEP08(2013)021}{{ JHEP} {\bfseries 1308}
  (2013) 021},
\href{http://arxiv.org/abs/1305.1001}{{\ttfamily arXiv:1305.1001}}.

\bibitem{Fukuyama:2012rw}
T.~Fukuyama, ``{SO(10) GUT in Four and Five Dimensions: A Review},''
  \href{http://dx.doi.org/10.1142/S0217751X13300081}{{ Int. J. Mod. Phys.}
  {\bfseries A28} (2013) 1330008},
\href{http://arxiv.org/abs/1212.3407}{{\ttfamily arXiv:1212.3407 [hep-ph]}}.

\bibitem{Mambrini:2015vna}
Y.~Mambrini, N.~Nagata, K.~A. Olive, J.~Quevillon, and J.~Zheng, ``{Dark Matter
  and Gauge Coupling Unification in Nonsupersymmetric SO(10) Grand Unified
  Models},'' \href{http://dx.doi.org/10.1103/PhysRevD.91.095010}{{ Phys. Rev.}
  {\bfseries D91} no.~9, (2015) 095010},
\href{http://arxiv.org/abs/1502.06929}{{\ttfamily arXiv:1502.06929 [hep-ph]}}.

\bibitem{Ellis:2018khn}
S.~A. Ellis, T.~Gherghetta, K.~Kaneta, and K.~A. Olive, ``{New Weak-Scale
  Physics from $SO(10)$ with High-Scale Supersymmetry},''
  \href{http://dx.doi.org/10.1103/PhysRevD.98.055009}{{ Phys. Rev. D}
  {\bfseries 98} no.~5, (2018) 055009},
  \href{http://arxiv.org/abs/1807.06488}{{\ttfamily arXiv:1807.06488
  [hep-ph]}}.

\bibitem{Ferrari:2018rey}
S.~Ferrari, T.~Hambye, J.~Heeck, and M.~H. Tytgat, ``{$SO(10)$ Paths to Dark
  Matter},'' \href{http://dx.doi.org/10.1103/PhysRevD.99.055032}{{ Phys. Rev.
  D} {\bfseries 99} no.~5, (2019) 055032},
  \href{http://arxiv.org/abs/1811.07910}{{\ttfamily arXiv:1811.07910
  [hep-ph]}}.

\bibitem{Chakrabortty:2019fov}
J.~Chakrabortty, R.~Maji, and S.~F. King, ``{Unification, Proton Decay and
  Topological Defects in non-SUSY GUTs with Thresholds},''
  \href{http://dx.doi.org/10.1103/PhysRevD.99.095008}{{ Phys. Rev.} {\bfseries
  D99} no.~9, (2019) 095008},
\href{http://arxiv.org/abs/1901.05867}{{\ttfamily arXiv:1901.05867 [hep-ph]}}.

\bibitem{Chakraborty:2019uxk}
M.~Chakraborty, M.~Parida, and B.~Sahoo, ``{Triplet Leptogenesis, Type-II
  Seesaw Dominance, Intrinsic Dark Matter, Vacuum Stability and Proton Decay in
  Minimal $SO(10)$ Breakings},''
  \href{http://dx.doi.org/10.1088/1475-7516/2020/01/049}{{ JCAP} {\bfseries 01}
  (2020) 049}, \href{http://arxiv.org/abs/1906.05601}{{\ttfamily
  arXiv:1906.05601 [hep-ph]}}.

\bibitem{Chang:1985zq}
D.~Chang and A.~Kumar, ``{Symmetry Breaking of SO(10) by 210-dimensional Higgs
  Boson and the Michel's Conjecture},''
  \href{http://dx.doi.org/10.1103/PhysRevD.33.2695}{{ Phys. Rev. D} {\bfseries
  33} (1986) 2695}.

\bibitem{McKay:1981}
W.~G. McKay and J.~Patera, { Tables of Dimensions, Indices, and Branching Rules
  for Representations of Simple Lie Algebras}.
\newblock Marcel Dekker, Inc., New York, 1981.

\bibitem{Fonseca:2011sy}
R.~M. Fonseca, ``{Calculating the Renormalisation Group Equations of a SUSY
  Model with Susyno},'' \href{http://dx.doi.org/10.1016/j.cpc.2012.05.017}{{
  Comput.Phys.Commun.} {\bfseries 183} (2012) 2298--2306},
\href{http://arxiv.org/abs/1106.5016}{{\ttfamily arXiv:1106.5016 [hep-ph]}}.

\bibitem{Feger:2012bs}
R.~Feger and T.~W. Kephart, ``{LieART - A Mathematica Application for Lie
  Algebras and Representation Theory},''
  \href{http://dx.doi.org/10.1016/j.cpc.2014.12.023}{{ Comput.Phys.Commun.}
  {\bfseries 192} (2015) 166--195},
\href{http://arxiv.org/abs/1206.6379}{{\ttfamily arXiv:1206.6379 [math-ph]}}.

\bibitem{Feger:2019tvk}
R.~Feger, T.~W. Kephart, and R.~J. Saskowski, ``{LieART 2.0 -- A Mathematica
  Application for Lie Algebras and Representation Theory},'' { Comput. Phys.
  Commun.} {\bfseries 257} (2020) 107490,
\href{http://arxiv.org/abs/1912.10969}{{\ttfamily arXiv:1912.10969 [hep-th]}}.

\bibitem{Fonseca:2020vke}
R.~M. Fonseca, ``{GroupMath: A Mathematica Package for Group Theory
  Calculations},'' \href{http://arxiv.org/abs/2011.01764}{{\ttfamily
  arXiv:2011.01764 [hep-th]}}.

\bibitem{Machacek:1983tz}
M.~E. Machacek and M.~T. Vaughn, ``{Two Loop Renormalization Group Equations in
  a General Quantum Field Theory. 1. Wave Function Renormalization},''
\href{http://dx.doi.org/10.1016/0550-3213(83)90610-7}{{ Nucl. Phys.} {\bfseries
  B222} (1983) 83}.

\bibitem{Machacek:1983fi}
M.~E. Machacek and M.~T. Vaughn, ``{Two Loop Renormalization Group Equations in
  a General Quantum Field Theory. 2. Yukawa Couplings},''
\href{http://dx.doi.org/10.1016/0550-3213(84)90533-9}{{ Nucl. Phys.} {\bfseries
  B236} (1984) 221}.

\bibitem{Machacek:1984zw}
M.~E. Machacek and M.~T. Vaughn, ``{Two Loop Renormalization Group Equations in
  a General Quantum Field Theory. 3. Scalar Quartic Couplings},''
\href{http://dx.doi.org/10.1016/0550-3213(85)90040-9}{{ Nucl. Phys.} {\bfseries
  B249} (1985) 70}.

\bibitem{Zyla:2020zbs}
{\bfseries Particle Data Group} Collaboration, P.~Zyla {et al.}, ``{Review of
  Particle Physics},'' \href{http://dx.doi.org/10.1093/ptep/ptaa104}{{ PTEP}
  {\bfseries 2020} no.~8, (2020) 083C01}.

\bibitem{Mohapatra2002}
R.~N. Mohapatra, { {Unification and Supersymmetry -The Frontiers of
  Quarks-Lepton Physics-}}.
\newblock Springer, 2002.

\bibitem{Deshpande:1992au}
N.~Deshpande, E.~Keith, and P.~B. Pal, ``{Implications of LEP Results for
  SO(10) Grand Unification},''
\href{http://dx.doi.org/10.1103/PhysRevD.46.2261}{{ Phys.Rev.} {\bfseries D46}
  (1992) 2261--2264}.

\bibitem{Deshpande:1992em}
N.~Deshpande, E.~Keith, and P.~B. Pal, ``{Implications of LEP Results for
  SO(10) Grand Unification with Two Intermediate Stages},''
  \href{http://dx.doi.org/10.1103/PhysRevD.47.2892}{{ Phys.Rev.} {\bfseries
  D47} (1993) 2892--2896},
\href{http://arxiv.org/abs/hep-ph/9211232}{{\ttfamily arXiv:hep-ph/9211232
  [hep-ph]}}.

\bibitem{Nath:2006ut}
P.~Nath and P.~Fileviez~Perez, ``{Proton Stability in Grand Unified Theories,
  in Strings and in Branes},''
  \href{http://dx.doi.org/10.1016/j.physrep.2007.02.010}{{ Phys. Rept.}
  {\bfseries 441} (2007) 191--317},
  \href{http://arxiv.org/abs/hep-ph/0601023}{{\ttfamily arXiv:hep-ph/0601023}}.

\bibitem{Takenaka:2020vqy}
{\bfseries Super-Kamiokande} Collaboration, A.~Takenaka {et al.}, ``{Search for
  Proton Decay via $p\to e^+\pi^0$ and $p\to \mu^+\pi^0$ with an Enlarged
  Fiducial Volume in Super-Kamiokande I-IV},''
  \href{http://dx.doi.org/10.1103/PhysRevD.102.112011}{{ Phys. Rev. D}
  {\bfseries 102} no.~11, (2020) 112011},
  \href{http://arxiv.org/abs/2010.16098}{{\ttfamily arXiv:2010.16098
  [hep-ex]}}.

\bibitem{Dorsner:2016wpm}
I.~Dor\v{s}ner, S.~Fajfer, A.~Greljo, J.~Kamenik, and N.~Ko\v{s}nik, ``{Physics
  of Leptoquarks in Precision Experiments and at Particle Colliders},''
  \href{http://dx.doi.org/10.1016/j.physrep.2016.06.001}{{ Phys. Rept.}
  {\bfseries 641} (2016) 1--68},
  \href{http://arxiv.org/abs/1603.04993}{{\ttfamily arXiv:1603.04993
  [hep-ph]}}.

\bibitem{Babu:2015bna}
K.~S. Babu and S.~Khan, ``{Minimal Nonsupersymmetric $SO(10)$ Model: Gauge
  Coupling Unification, Proton Decay, and Fermion Masses},''
  \href{http://dx.doi.org/10.1103/PhysRevD.92.075018}{{ Phys. Rev. D}
  {\bfseries 92} no.~7, (2015) 075018},
  \href{http://arxiv.org/abs/1507.06712}{{\ttfamily arXiv:1507.06712
  [hep-ph]}}.

\bibitem{Chang:1983fu}
D.~Chang, R.~N. Mohapatra, and M.~K. Parida, ``{Decoupling Parity and $SU(2)_R$
  Breaking Scales: A New Approach to Left-Right Symmetric Models},''
  \href{http://dx.doi.org/10.1103/PhysRevLett.52.1072}{{ Phys. Rev. Lett.}
  {\bfseries 52} (1984) 1072}.

\bibitem{Chang:1984uy}
D.~Chang, R.~N. Mohapatra, and M.~K. Parida, ``{A New Approach to Left-Right
  Symmetry Breaking in Unified Gauge Theories},''
  \href{http://dx.doi.org/10.1103/PhysRevD.30.1052}{{ Phys. Rev. D} {\bfseries
  30} (1984) 1052}.

\bibitem{Jones:1981we}
D.~R.~T. Jones, ``{The Two Loop beta Function for a $G_1\times G_2$ Gauge
  Theory},'' \href{http://dx.doi.org/10.1103/PhysRevD.25.581}{{ Phys. Rev. D}
  {\bfseries 25} (1982) 581}.

\bibitem{Hall:1980kf}
L.~J. Hall, ``{Grand Unification of Effective Gauge Theories},''
\href{http://dx.doi.org/10.1016/0550-3213(81)90498-3}{{ Nucl. Phys.} {\bfseries
  B178} (1981) 75--124}.

\bibitem{Chang:1984qr}
D.~Chang, R.~N. Mohapatra, J.~Gipson, R.~E. Marshak, and M.~K. Parida,
  ``{Experimental Tests of New $SO(10)$ Grand Unification},''
  \href{http://dx.doi.org/10.1103/PhysRevD.31.1718}{{ Phys. Rev. D} {\bfseries
  31} (1985) 1718}.

\bibitem{Baring:2015sza}
M.~G. Baring, T.~Ghosh, F.~S. Queiroz, and K.~Sinha, ``{New Limits on the Dark
  Matter Lifetime from Dwarf Spheroidal Galaxies Using Fermi-LAT},''
  \href{http://dx.doi.org/10.1103/PhysRevD.93.103009}{{ Phys. Rev. D}
  {\bfseries 93} no.~10, (2016) 103009},
  \href{http://arxiv.org/abs/1510.00389}{{\ttfamily arXiv:1510.00389
  [hep-ph]}}.

\bibitem{PalomaresRuiz:2007ry}
S.~Palomares-Ruiz, ``{Model-Independent Bound on the Dark Matter Lifetime},''
  \href{http://dx.doi.org/10.1016/j.physletb.2008.05.040}{{ Phys. Lett. B}
  {\bfseries 665} (2008) 50--53},
  \href{http://arxiv.org/abs/0712.1937}{{\ttfamily arXiv:0712.1937
  [astro-ph]}}.

\bibitem{Covi:2009xn}
L.~Covi, M.~Grefe, A.~Ibarra, and D.~Tran, ``{Neutrino Signals from Dark Matter
  Decay},'' \href{http://dx.doi.org/10.1088/1475-7516/2010/04/017}{{ JCAP}
  {\bfseries 04} (2010) 017}, \href{http://arxiv.org/abs/0912.3521}{{\ttfamily
  arXiv:0912.3521 [hep-ph]}}.

\bibitem{Belyaev:2012qa}
A.~Belyaev, N.~D. Christensen, and A.~Pukhov, ``{CalcHEP 3.4 for Collider
  Physics Within and Beyond the Standard Model},''
  \href{http://dx.doi.org/10.1016/j.cpc.2013.01.014}{{ Comput. Phys. Commun.}
  {\bfseries 184} (2013) 1729--1769},
  \href{http://arxiv.org/abs/1207.6082}{{\ttfamily arXiv:1207.6082 [hep-ph]}}.

\bibitem{Belanger:2018ccd}
G.~B\'elanger, F.~Boudjema, A.~Goudelis, A.~Pukhov, and B.~Zaldivar,
  ``{micrOMEGAs5.0 : Freeze-in},''
  \href{http://dx.doi.org/10.1016/j.cpc.2018.04.027}{{ Comput. Phys. Commun.}
  {\bfseries 231} (2018) 173--186},
  \href{http://arxiv.org/abs/1801.03509}{{\ttfamily arXiv:1801.03509
  [hep-ph]}}.

\bibitem{Sirunyan:2018owy}
{\bfseries CMS} Collaboration, A.~M. Sirunyan {et al.}, ``{Search for Invisible
  Decays of a Higgs Boson Produced Through Vector Boson Fusion in Proton-Proton
  Collisions at $\sqrt{s} =$ 13 TeV},''
  \href{http://dx.doi.org/10.1016/j.physletb.2019.04.025}{{ Phys. Lett. B}
  {\bfseries 793} (2019) 520--551},
  \href{http://arxiv.org/abs/1809.05937}{{\ttfamily arXiv:1809.05937
  [hep-ex]}}.

\bibitem{Aaboud:2019rtt}
{\bfseries ATLAS} Collaboration, M.~Aaboud {et al.}, ``{Combination of Searches
  for invisible Higgs Boson Decays with the ATLAS Experiment},''
  \href{http://dx.doi.org/10.1103/PhysRevLett.122.231801}{{ Phys. Rev. Lett.}
  {\bfseries 122} no.~23, (2019) 231801},
  \href{http://arxiv.org/abs/1904.05105}{{\ttfamily arXiv:1904.05105
  [hep-ex]}}.

\bibitem{Chen:2014ask}
C.-Y. Chen, S.~Dawson, and I.~M. Lewis, ``{Exploring Resonant Di-Higgs Boson
  Production in the Higgs Singlet Model},''
  \href{http://dx.doi.org/10.1103/PhysRevD.91.035015}{{ Phys. Rev. D}
  {\bfseries 91} no.~3, (2015) 035015},
  \href{http://arxiv.org/abs/1410.5488}{{\ttfamily arXiv:1410.5488 [hep-ph]}}.

\bibitem{Fermi-LAT:2016uux}
{\bfseries Fermi-LAT, DES} Collaboration, A.~Albert {et al.}, ``{Searching for
  Dark Matter Annihilation in Recently Discovered Milky Way Satellites with
  Fermi-LAT},'' \href{http://dx.doi.org/10.3847/1538-4357/834/2/110}{{
  Astrophys. J.} {\bfseries 834} no.~2, (2017) 110},
  \href{http://arxiv.org/abs/1611.03184}{{\ttfamily arXiv:1611.03184
  [astro-ph.HE]}}.

\bibitem{Binder:2017rgn}
T.~Binder, T.~Bringmann, M.~Gustafsson, and A.~Hryczuk, ``{Early Kinetic
  Decoupling of Dark Matter: When the Standard Way of Calculating the Thermal
  Relic Density Fails},'' \href{http://dx.doi.org/10.1103/PhysRevD.96.115010}{{
  Phys. Rev. D} {\bfseries 96} no.~11, (2017) 115010},
  \href{http://arxiv.org/abs/1706.07433}{{\ttfamily arXiv:1706.07433
  [astro-ph.CO]}}. [Erratum: Phys.Rev.D 101, 099901 (2020)].

\bibitem{Abe:2020obo}
T.~Abe, ``{Effect of the Early Kinetic Decoupling in a Fermionic Dark Matter
  Model},'' \href{http://dx.doi.org/10.1103/PhysRevD.102.035018}{{ Phys. Rev.
  D} {\bfseries 102} no.~3, (2020) 035018},
  \href{http://arxiv.org/abs/2004.10041}{{\ttfamily arXiv:2004.10041
  [hep-ph]}}.

\end{thebibliography}\endgroup

\end{document}